\documentclass{ametsocV5}
\nolinenumbers

\usepackage{amsmath,amsfonts,amssymb,bm}
\usepackage{mathptmx}
\usepackage{newtxtext}
\usepackage{newtxmath}

\title{Quantifying the Predictability of ENSO Complexity Using a Statistically Accurate Multiscale Stochastic Model and Information Theory}
\authors{Xianghui Fang$^{1,2,3}$ and Nan Chen$^4$\correspondingauthor{Nan Chen, chennan@math.wisc.edu}}

\affiliation{1. Department of Atmospheric and Oceanic Sciences and Institute of Atmospheric Sciences, Fudan University, 220 Handan Rd., Shanghai 200433, China\\%
2. Innovation Center of Ocean and Atmosphere System, Zhuhai Fudan Innovation Research Institute, Zhuhai 518057, China\\%
3. Shanghai Scientific Frontier Base of Air-Sea Interaction, Shanghai 200438, China\\%
4. Department of Mathematics, University of Wisconsin-Madison, Madison, WI, USA}

\abstract{An information theory based framework is developed to assess the predictability of the ENSO complexity, which includes different types of the ENSO events with diverse characteristics in spatial patterns, peak intensities and temporal evolutions. The information theory advances a unique way to quantify the forecast uncertainty and allows to distinguish the predictability limit of each type of event. With the assistance of a recently developed multiscale stochastic conceptual model that succeeds in capturing both the large-scale dynamics and many crucial statistical properties of the observed ENSO complexity, it is shown that different ENSO events possess very distinct predictability limits. Beyond the ensemble mean value, the spread of the ensemble members also has remarkable contributions to the predictability. Specifically, while the result indicates that predicting the onset of the eastern Pacific (EP) El Ni\~nos is challenging, it reveals a universal tendency to convert strong predictability to skillful forecast for predicting many central Pacific (CP) El Ni\~nos about two years in advance. In addition, strong predictability is found for the La Ni\~na events, corresponding to the effectiveness of the El Ni\~no to La Ni\~na transitions. In the climate change scenario with the strengthening of the background Walker circulation, the predictability of sea surface temperature in the CP region has a significant response with a notable increase in summer and fall. Finally, the Gaussian approximation exhibits to be accurate in computing the information gain, which facilitates the use of more sophisticated models to study the ENSO predictability.}

\begin{document}

\maketitle


\section{Introduction}

El Ni\~no-Southern Oscillation (ENSO) is the most significant interannual climate signal in the tropics \citep{philander1983nino, ropelewski1987global, klein1999remote, mcphaden2006enso}. In the classical viewpoint, ENSO was often regarded as a phenomenon with symmetric and cyclical attributes \citep{jin1997equatorial}, in which the positive and negative phases are El Ni\~no and La Ni\~na, respectively. Yet, ENSO is known to show a significant diversity and irregularity \citep{capotondi2015understanding,timmermann2018nino}.
Many studies have suggested that there are at least two types of ENSO \citep{ashok2007nino, yu2007decadal, kao2009contrasting}. Based on the features during the mature phase, they are named as the eastern Pacific (EP) and the central Pacific (CP) types when the largest sea surface temperature (SST) anomaly is located near the coast of the South America and the dateline region, respectively \citep{yu2007decadal, kao2009contrasting}. In addition to spatial patterns, different ENSO events also exhibit diverse characteristics in peak intensities and temporal evolutions, which is known as the ENSO complexity \citep{hayashi2017enso, timmermann2018nino, boucharel2021influence}.

Due to its crucial impacts on both the regional and global climate, simulating and forecasting ENSO has long been a focus of research in related fields. Through unremitting efforts, considerable achievements have been made and a series of ENSO models with a hierarchy of complexity have been developed, including conceptual or box models \citep{schopf1988vacillations,picaut1996mechanism,jin1997equatorial,wang2004understanding}, simple models \citep{leeuwen2002balanced,fang2018simulating}, intermediate complex models (ICMs) \citep{zebiak1987model, battisti1989interannual, neelin1993modes,zhang2003new,thual2016simple,chen2022simple} and coupled general circulation models (CGCMs) (Planton et al., 2021). These models can generally successfully forecast the Ni\~no3.4 ($170^o$W-$120^o$W, $5^o$S-$5^o$N) SST index 6 to 12 months in advance, thereby providing a reliable basis for many other climatic predictions that depend on the forecast of ENSO \citep{jin2008current,barnston2012skill}.

In spite of these achievements, there is still an expectation to further improve the ENSO forecast skill. However, before that can be done, it is essential to quantitatively understand what is the upper limit of the forecast skill, i.e., the potential predictability of ENSO. Note that the forecast skill is tightly related to the model performance on depicting the physical processes, as well as the observational network, the initialization methods, specific prediction procedures and post-processing, etc., i.e., it connects a system or project \citep{tang2018progress}. However, the predictability is to measure the attribute of ENSO, which reflects its inherent characteristic rather than our ability to make practical prediction \citep{l2020enso}. Under the ``ideal model" framework, one can measure the growth dynamics of the prediction error and estimate the potential predictability by perturbing the initial conditions with a well designed or a series of random noises that have the amplitude much smaller than the accuracy of the observational system. Furthermore, in addition to the overall measurements, it is of more practical importance to explore the predictability of different types of the ENSO events due to their distinct features and impacts. Understanding the gap between the potential predictability and the prediction skill can provide guidelines to the improvement of the existing models and the current forecast methods. Such a study may also suggest that the efforts for further improving the forecast of certain types of the ENSO events is futile if they are inherently unpredictable.

In this paper, an information theory based framework \citep{delsole2004predictability, majda2005information,kleeman2011information} is developed to study the potential predictability of the ENSO complexity. Fundamentally different from the time series based deterministic methods, the information theory based framework exploits the statistical forecast to quantify the predictability of each individual event in a complex chaotic system. The indicator of the predictability within this framework is the time evolution of the information distance between the forecast distribution resulting from an ensemble forecast and the climatological one from the observational data. It denotes the additional information provided by the ensemble forecast that is beyond the prior known climatology, which is also called the information gain.
Therefore, the amplitude of the information distance suggests a natural measurement of the potential predictability.

There are several desirable features of adopting such a framework to understand the predictability of chaotic systems \citep{cover1999elements,kleeman2002measuring, delsole2007predictability}, including the ENSO. First, as its focus is the entire distribution, which is also called the probability density function (PDF), the quantification of forecast uncertainty has been taken into account in the assessment of the predictability. This is vital for chaotic systems as each single forecast trajectory often deviates quickly from the truth. Second, in contrast to the time series-based methods that require a large number of the ensemble mean forecast results at a fixed lead time with different initializations, the information theory allows to quantify the predictability of a specific event starting from a given initial condition. In fact, the forecast distribution at a fixed lead time can be formulated by the collection of different forecast ensemble members. Such a unique feature allows us to distinguish the predictability limit of different ENSO events, which is crucial for understanding and advancing the prediction of the ENSO complexity.

Although the concept of information theory has been proposed for decades, there exists a major barrier in applying it to effectually quantify the predictability of many complex natural phenomena. In fact, the effectiveness of the resulting predictability from the information theory requires the model that is used to faithfully represent the essential non-Gaussian statistics of ENSO. Otherwise, the model error rather than the intrinsic predictability may become the dominant component of the information distance \citep{giannakis2012quantifying, branicki2013non, chen2016model}. However, reproducing the statistical features is often not the most primary focus of the existing dynamical models, especially the ICMs and the CGCMs. Therefore, special cautions need to be paid when these models are incorporated into the information theory based framework. Although the information theory has been applied to study the overall El Ni\~no predictability in the EP region \citep{tang2005reliability} and leads to some interesting insights, it is unclear about the proportion of the explainable information gain there related to the actual predictability and due to the model error, respectively.
On the other hand, some simple purely data-driven or statistical models, such as the linear regression models and the linear inverse models, are able to recover certain basic statistics, for example, the climatology covariance \citep{dominiak2005improvement, alexander2008forecasting}. However, the lack of the crucial nonlinear dynamics and explainable physics may also impede these models for characterizing the predictability of nature since the transient behavior can be very different from that of the actual nonlinear chaotic systems.

To facilitate the quantification of the ENSO predictability within the information theory based framework, a recently developed conceptual multiscale stochastic model for the ENSO complexity is utilized in this study \citep{chen2022multiscale}. Although many simple or ICMs can reproduce most ENSO properties, to our knowledge, this is the first conceptual model that can accurately reproduce the ENSO diversity and many important statistical features of the ENSO complexity. The model starts with a deterministic three-region system for the interannual variabilities. Then two stochastic processes of the intraseasonal and decadal variation are incorporated. The model can reproduce not only the general properties of the observed ENSO events, but also the complexity in patterns (e.g., CP vs. EP El Ni\~no events), intensities (e.g., 10-20 year reoccurrence of extreme El Ni\~nos), and temporal evolutions (e.g., more multi-year La Ni\~nas than multi-year El Ni\~nos). Besides, the model can also accurately simulate the PDFs, the power spectra and the seasonal phase-locking of the SST variations in both the CP and EP regions. A brief description of the model will be introduced in the following Section, while more detailed descriptions and experiment tests can be found in the original paper.
In addition, the low computational cost of such a conceptual model allows us to use a large number of the ensembles for the study here, avoiding the statistical error due to the undersampling issue. Therefore, this model is a suitable candidate for us to explore the predictability of ENSO complexity from the perspective of the above mentioned information theory based framework. Notably, understanding the predictability of the ENSO complexity is a much more challenging but important task than studying the overall SST predictability in only the EP region.

The rest for this paper is organized as follows. The general framework of quantifying the predictability of complex natural phenomena using information theory is presented in Section \ref{Sec:Information_Theory}. Section \ref{Sec:Data_Model} introduces the datasets used in this study. It also includes a brief summary of the statistically accurate multiscale stochastic conceptual model. Section \ref{Sec:Overall} presents the overall ENSO predictability while the predictability of different ENSO events related to the ENSO complexity is analyzed in Section \ref{Sec:Complexity}. The sensitivity analysis, including the ENSO predictability in the climate change scenarios and the role of each multiscale component in affecting the ENSO prediction, is investigated in Section \ref{Sec:Sensitivity}. The conclusions and discussions are provided in Section \ref{Sec:Conclusions}.

\section{An Information Theory based Framework: Quantifying the Predictability of Complex Natural Phenomena}\label{Sec:Information_Theory}
\begin{figure}[h]
\centerline{\includegraphics[width=\textwidth]{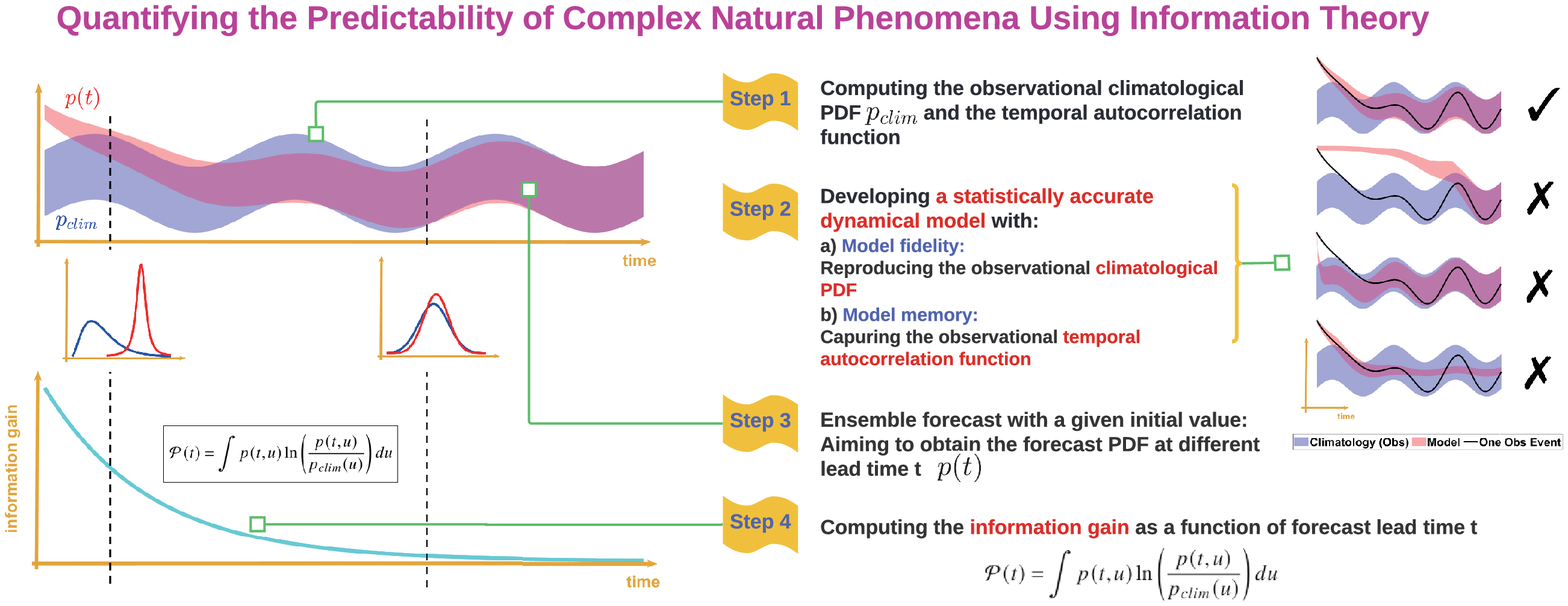}}
\caption{Schematic illustration of quantifying the predictability of complex natural phenomena using information theory.} \label{ENSO_Skematic}
\end{figure}

We start with building the general framework of using information theory to quantify the predictability of complex natural phenomena, which will then be applied to the study of the ENSO complexity \citep{tang2005reliability}. The key quantities within the framework are a) the forecast PDF $p(t,u)$ from a suitable model with $t$ and $u$ being the time and the specific variable, respectively; b) the initial value and climatological PDF $p_{clim}$ obtained from the observational data; c) a simple and effective approach to compute the information distance between $p(t,u)$ and $p_{clim}$. More precise definitions of $p(t,u)$ and $p_{clim}$ will be provided below. A schematic illustration of the framework is summarized in Figure \ref{ENSO_Skematic}, which contains four steps.\medskip

\noindent\emph{Step 1. Computing the climatological PDF $p_{clim}$ and the temporal autocorrelation function (ACF) from observational data.}

The observational climatological PDF $p_{clim}$ plays a vital role, serving as a reference solution, in the study of the predictability via the information theory based framework.
Given the observational data, say for simplicity an index (time series), the climatological PDF $p_{clim}$ can be computed by first drawing the histogram from the data and then normalizing it. In practice, $p_{clim}$ is often a time periodic function with period being one year, representing the seasonal cycle. Therefore, if a partition consisting of 12 equidistant points is used to formulate such a time-periodic function, then all the data corresponding to a specific month can be used to compute the associated climatological PDF for that month.

On the other hand, the ACF is not explicitly involved in computing the information gain for quantifying the predictability. Nevertheless, it is of fundamental importance in Step 2, serving as one of the central guidelines for the development of a statistically accurate dynamical model that is used to compute the time evolution of the forecast PDF $p(t)$. Autocorrelation is the correlation of a signal with a delayed copy of itself as a function of delay \citep{gardiner2009stochastic}. The ACF measures the overall memory of a chaotic system and describes the averaged convergence rate of the statistics towards the climatology. For a zero mean and stationary scalar time series $u$, which can be a specific dimension of a vector state variable or simply an index, the ACF can be calculated as
\begin{equation}\label{Definition_ACF}
  \mbox{ACF}(t) = \lim_{T\to\infty}\frac{1}{T}\int_0^T\frac{u(t+t')u^*(t')}{\mbox{Var}(u)}dt'.
\end{equation}
where $t$ is the delay and $T$ is the total length of the time series. The limit is taken for the mathematical rigor. In practice, the ACF is approximated by using a finite value of $T$ in \eqref{Definition_ACF}, provided that $T$ is sufficiently large.  \medskip

\noindent\emph{Step 2. Development of a statistically accurate dynamical model.}

This is one of the most vital and challenging steps that involve the incorporation of important physics into the framework. As mentioned above, the calculated information gain is exact only when the perfect model of nature is adopted. Otherwise, the resulting information gain used to explain the predictability can be polluted by the model error. In practice, two necessary conditions have to be satisfied to mitigate the model error in calculating the information gain \citep{majda2018model, majda2018strategies}.

\noindent\emph{(a) Model fidelity: The model must have the skill to reproduce the climatological PDF of nature.} Such a necessary condition guarantees the consistency of the model at long lead times with the observational statistics. If the model lacks the fidelity, then the information distance between $p(t,u)$ and $p_{clim}$ at long lead time will never become zero. As a consequence, the predictability limit will become infinity, which is erroneous.

\noindent\emph{(b) Model memory: The model must be able to capture the overall temporal autocorrelation of nature.} Satisfying the model fidelity guarantees the long-term statistics being captured. However, without additional constraints, the model fidelity itself is not sufficient to ensure that the time evolution of the statistics and the associated relaxation tendency towards the climatological distribution of the model are consistent with those of nature. In other words, the time evolution of the information gain computed from the model can be biased due to the failure of model in capturing the transient behavior of nature. Since the ACF measures the overall memory of a chaotic system, the difference between the temporal ACFs can be utilized as a simple and effective criterion to characterize the similarity of the transient behavior between the two systems. A suitable model that can be used for the information theory based framework should have the ACF that resembles the one of nature.

To provide intuitions, some illustrative example of the model fidelity and the model memory are included at the right top corner of Figure \ref{ENSO_Skematic}. The blue shading area represents the time evolution of the climatological PDF computed from observational data. The red shading area stands for the time evolution of the model PDF starting from a specific observed event, the actual time evolution of which is denoted by the black solid curve. The four rows show four different examples of model behaviours. The model in the first row satisfies both the model fidelity and model memory since the time evolution of the model statistics tracks the observational trajectory and it converges to the same climatology as the observations. The models in the second and the third row also satisfy the model fidelity but the relaxation of the statistics towards the climatology are either much slower or much faster than that of nature. These two models fail to characterize the overall memory of nature. The model in the last row has a consistent autocorrelation with nature but the variance of the climatological PDF is underestimated compared with nature, which means the model fidelity is not satisfied.\medskip

\noindent\emph{Step 3. Ensemble forecast with a given initial value: aiming to obtain the forecast PDF $p(t)$ at different lead time $t$.}

The forecast PDF $p(t)$ is computed by running the model repeatedly forward in time starting from a given initial condition from observations (with a slight perturbation if needed). Due to the random forcing or the intrinsic chaotic behavior of the model, different realizations will be distinct with each other. Collecting all the forecast realizations allows to form a forecast PDF $p(t)$ at each lead time $t$.\medskip

\noindent\emph{Step 4. Computing the information gain as a function of forecast lead time $t$.}

With the observational climatological PDF $p_{clim}$ and the time evolution of the forecast PDF $p(t)$ in hand from the previous steps, what remains is to develop an appropriate information criterion to compute the difference between $p(t)$ and $p_{clim}$. Such a difference represents the additional information in $p(t)$ beyond the observational climatology, which naturally reflects the potential predictability \citep{kleeman2011information, buizza2015forecast}.

Since the comparison is two PDFs, the information theory is a more appropriate choice than the time series-based measurements. One natural way to assess the information gain in $p(t)$ compared with the climatological PDF $p_{clim}$ is through the relative entropy $\mathcal{P}(p(t),p_{clim})$  \citep{majda2005information,delsole2007predictability, kleeman2011information},
\begin{equation}\label{Relative_Entropy}
  \mathcal{P}(p(t),p_{clim}) = \int p(t,u)\log\left(\frac{p(t,u)}{p_{clim}(u)}\right)du,
\end{equation}
where $u$ is the state variable that is integrated out. The relative entropy is also known as the Kullback-Leibler divergence or information divergence \citep{kullback1951information, kullback1959statistics, kullback1987letter}.
Despite the lack of symmetry, the relative entropy has two attractive features. First, $\mathcal{P}(p(t),p_{clim}) \geq 0$ with equality if and only if $p(t)=p_{clim}$. Second, $\mathcal{P}(p(t),p_{clim})$ is invariant under general nonlinear changes of variables \citep{kleeman2011information,majda2018model}. These provide an attractive framework for assessing the information gain as well as quantifying the model error and model sensitivity in other applications \citep{majda2002mathematical, kleeman2002measuring, delsole2005predictability,tang2007predictability, branstator2010two,teng2011initial, giannakis2012quantifying, chen2014information, liu2017predictability}.

One practical setup for utilizing the framework of information theory in many applications arises when both the measurements involve only the mean and covariance so that
\begin{equation*}
p(t)\approx p^G(t)\sim\mathcal{N}(\bar{u}, R)\qquad \mbox{and}\qquad p_{clim}\approx p^G_{clim}\sim\mathcal{N}(\bar{u}_{clim}, R_{clim})
\end{equation*}
can be approximated by Gaussian distributions. In this case, $\mathcal{P}(p(t),p_{clim})$ has the explicit formula
\begin{equation}\label{Signal_Dispersion}
  \mathcal{P}(p(t),p_{clim}) = \left[\frac{1}{2}(\bar{u}-\bar{u}_{clim})^2R_{clim}^{-1}\right] + \left[-\frac{1}{2}\log(RR_{clim}^{-1}) + \frac{1}{2}(RR_{clim}^{-1}-1)\right].
\end{equation}
In \eqref{Signal_Dispersion}, the first term in brackets is called `signal', reflecting the information difference in the mean but weighted by the inverse of the climatological variance, $R_{clim}$, whereas the second term in brackets, called `dispersion', involves only the information distance regarding the covariance ratio, $RR_{clim}^{-1}$. The signal and dispersion terms in \eqref{Signal_Dispersion} are individually invariant under any (linear) change of variables which maps Gaussian distributions to Gaussians.

Note that, although the relative entropy has a lower bound $0$ when $p(t)=p_{clim}$, it has no upper bound. For the convenience of presentation, a rescaled version of the relative entropy will be utilized in the remaining of the paper \citep{giannakis2012information},
\begin{equation}\label{Relative_Entropy_Rescaled}
  \mathcal{E} = 1-\exp(-\mathcal{P})
\end{equation}
and therefore $\mathcal{E}$ is scaled to the interval $[0, 1]$. The rescaled relative entropy $\mathcal{E}$ is still a monotonically increasing function as the difference between $p(t)$ and $p_{clim}$. In the following, the relative entropy, together with its signal and dispersion components, refers to the rescaled version in \eqref{Relative_Entropy_Rescaled}, which is served as the indicator of the predictability.

\section{Observational Datasets and the Statistically Accurate Model for the ENSO Complexity}\label{Sec:Data_Model}
\subsection{Data}
The monthly ocean temperature and current data are both from the GODAS reanalysis dataset \citep{behringer2004evaluation}. The thermocline depth along the equatorial Pacific is approximated from the potential temperature as the depth of the $20^o$C isotherm. The analysis period is from 1982 to 2019. Anomalies presented in this study are calculated by removing the monthly mean climatology of the whole period. In this work, the Ni\~no4 ($T_C$) and Ni\~no3 ($T_E$) indices are the average of SST anomalies over the regions $160^o$E-$150^o$W, $5^o$S-$5^o$N (CP) and $150^o$W-$90^o$W, $5^o$S-$5^o$N (EP), respectively. The $h_W$ index is the mean thermocline depth anomaly over the western Pacific (WP) region ($120^o$E-$180^o$, $5^o$S-$5^o$N) while the $u$ index is the mean mixed-layer zonal current in the CP region.

The daily zonal wind data at 850 hPa from the NCEP–NCAR reanalysis \citep{kalnay1996ncep} is used to depict the intraseasonal wind bursts. By removing the daily mean climatology, the wind burst index is obtained by averaging the anomalies over the WP region. Besides, the Walker circulation strength index is adopted to measure the effect of the decadal variation in the characteristics of ENSO. Based on the definition of \citet{kang2020walker}, it is defined as the sea level pressure difference over the CP/EP ($160^o$W-$80^o$W, $5^o$S-$5^o$N) and over the Indian Ocean/WP ($80^o$E-$160^o$E, $5^o$S-$5^o$N). Note that the monthly zonal SST gradient between the WP and CP region is highly correlated with this Walker circulation strength index (correlation coefficient of $\sim0.85$), suggesting the significant air–sea interaction over the equatorial Pacific. Since the latter is more directly related to the zonal advective feedback strength over the CP region, the decadal model ($I$) mainly illustrates this variable.

\subsection{Definitions of different types of the ENSO events}\label{Subsec:Def_Events}
To quantify the complexity of ENSO, the definitions of different El Ni\~no and La Ni\~na events are as follows [based on the average SST anomalies during boreal winter (December–January–February; DJF)]: Following the definition in \citet{kug2009two}, when the EP is warmer than the CP and is greater than $0.5^o$C, it is classified as the EP El Ni\~no. Among this, based on the definitions used by \citet{wang2019historical}, an extreme El Ni\~no event corresponds to the situation that the maximum of EP SST anomaly from April to the next March is larger than $2.5^o$C. When the CP is warmer than the EP and is larger than $0.5^o$C, the event is then defined as a CP El Ni\~no. Finally, when either the CP and EP SST anomaly is cooler than $-0.5^o$C, it is defined as a La Ni\~na event.
Note that a more thorough definition for the two types of ENSO has been introduced in \citet{takahashi2011enso}, in which the indies of E and C that are transformed from Ni\~no4 and Ni\~no1+2 are used. However, since only the physical properties of the Ni\~no3 and Ni\~no4 indices are explicitly depicted in our conceptual model. Using the E and C measurements will bring inevitable biases. As a result, we just used the simplest definition on categorizing the ENSO types based the $T_E$ and $T_C$ indices.

\subsection{The multiscale stochastic model for ENSO complexity}
In this work, a multiscale stochastic conceptual model is used to study the predictability of ENSO complexity, which was recently developed in \cite{chen2022multiscale}. This is a three-region model, aiming to reproduce the observed dynamical and statistical features in both the CP and EP regions. The combination of suitable stochastic parameterizations and nonlinearities allows the model to generate different types of the observed ENSO events and capture the associated statistics. Here, we briefly summarize the main components of this model, while more details can be found in the original paper.

The model starts with a deterministic, linear and stable system for the interannual variabilities \citep{fang2018three}. It is a general extension of the classical recharge oscillator model \citep{jin1997equatorial} and depicts the air-sea interactions over the entire WP, CP and EP. That is, it includes both the ocean heat content discharge/recharge and the ocean zonal advection. Then, two stochastic processes with multiplicative noise describing the intraseasonal wind bursts and the decadal variation in the Walker circulation are incorporated to depict ENSO’s irregularity and the decadal variation in the strength and occurrence frequency of different ENSO events \citep{timmermann2018nino,fang2020brief}. The model reads:

	\begin{subequations}\label{Model_Stochastic}
		\begin{align}
			\frac{d u}{d t} &= -r u - \frac{\alpha_1 b_0 \mu}{2} (T_C + T_E) + {\beta_u\tau} + \sigma_u\dot{W}_u,\label{Model_Stochastic_u}\\
			\frac{d h_W}{d t} &= -r h_W - \frac{\alpha_2 b_0 \mu}{2} (T_C + T_E) + {\beta_h\tau} + \sigma_h\dot{W}_h,\label{Model_Stochastic_h_W}\\
			\frac{d T_C}{d t} &= \left(\frac{\gamma b_0\mu}{2} - c_1(T_C)\right) T_C + \frac{\gamma b_0 \mu}{2} T_E + \gamma h_W + \sigma u + C_u + \beta_C\tau +  \sigma_C\dot{W}_C,\label{Model_Stochastic_T_C}\\
			\frac{d T_E}{d t} &= \gamma h_W + \left(\frac{3\gamma b_0 \mu}{2} - c_2\right) T_E - \frac{\gamma b_0 \mu}{2} T_C + {\beta_E\tau}  + \sigma_E\dot{W}_E\label{Model_Stochastic_T_E}\\
			{\frac{d \tau}{d t}} & {= -d_\tau \tau + \sigma_\tau(T_C) \dot{W}_\tau},\label{Model_Stochastic_tau}\\
			{\frac{d I}{d t}} & {= -\lambda (I - m) + \sigma_I(I) \dot{W}_I}.\label{Model_Stochastic_I}
		\end{align}
	\end{subequations}
Here, the interannual component, i.e., Eqs. \eqref{Model_Stochastic_u}-\eqref{Model_Stochastic_T_E}, depicts the main dynamics for the ENSO; the intraseasonal equation \eqref{Model_Stochastic_tau} represents the amplitude of the random wind bursts ($\tau$); and the decadal part \eqref{Model_Stochastic_I} represents the variation in the strength of the background Walker circulation ($I$). In the model, $T_C$ and $T_E$ are the SST in the CP and EP, $u$ is the ocean zonal current in the CP and $h_W$ is the thermocline depth in the WP. As was discussed in \cite{chen2022multiscale}, $I$ also stands for the zonal SST difference between the WP and CP, which directly determines the strength of the zonal advective feedback. The parameter $m$ is
the mean of $I$, which can be computed from its PDF.

In this model, the stochasticity plays a crucial role in coupling variables at different time scales and parameterizing the unresolved features in the model. First, the intraseasonal variation $\tau$ is depicted by a simple stochastic differential equation with a state-dependent (i.e., multiplicative) noise coefficient $\sigma_\tau$, where $\dot{W}_\tau$ is a white noise source. $\tau$ is then coupled to the processes of the interannual part serving as external forcings. In addition, four Gaussian random noises $\sigma_u\dot{W}_u$, $\sigma_h\dot{W}_h$, $\sigma_C\dot{W}_C$ and $\sigma_E\dot{W}_E$ are further added to the processes describing the interannual variabilities, which effectively parameterize the additional contributions that are not explicitly modeled, such as the subtropical atmospheric forcing at the Pacific Ocean and the influences from the other Ocean basins. In a more general sense, these stochastic noises can be regarded as the simplest way for stochastic parameterization, which increases the model variability such that the PDFs of the model variables can better match those of the observational data \citep{palmer2009stochastic}. Second, since the details of the background Walker circulation consist of uncertainties and randomness \citep{chen2015strong}, a simple but effective stochastic process is used to describe the temporal evolution of the decadal variability $I$ \citep{yang2021enso}, where $\dot{W}_I$ in Eq. \eqref{Model_Stochastic_I} is another white noise source. The multiplicative noise in the process of $I$ is aimed at guaranteeing
its positivity due to the fact that the long-term average of the background Walker circulation is non-negative.
Besides, the effects of seasonality are added to both the wind activity and the collective damping to depict the seasonal phase-locking characteristics realistically, which manifests as the tendency of ENSO events to peak during boreal winter \citep{tziperman1997mechanisms, stein2014enso,fang2021effect}.

The dimensional units and the parameters in the coupled model are summarized in Table \ref{Table_Parameters}.

\subsection{The dynamical and statistical features of the model}
\begin{figure}[h]
\centerline{\includegraphics[width=\textwidth]{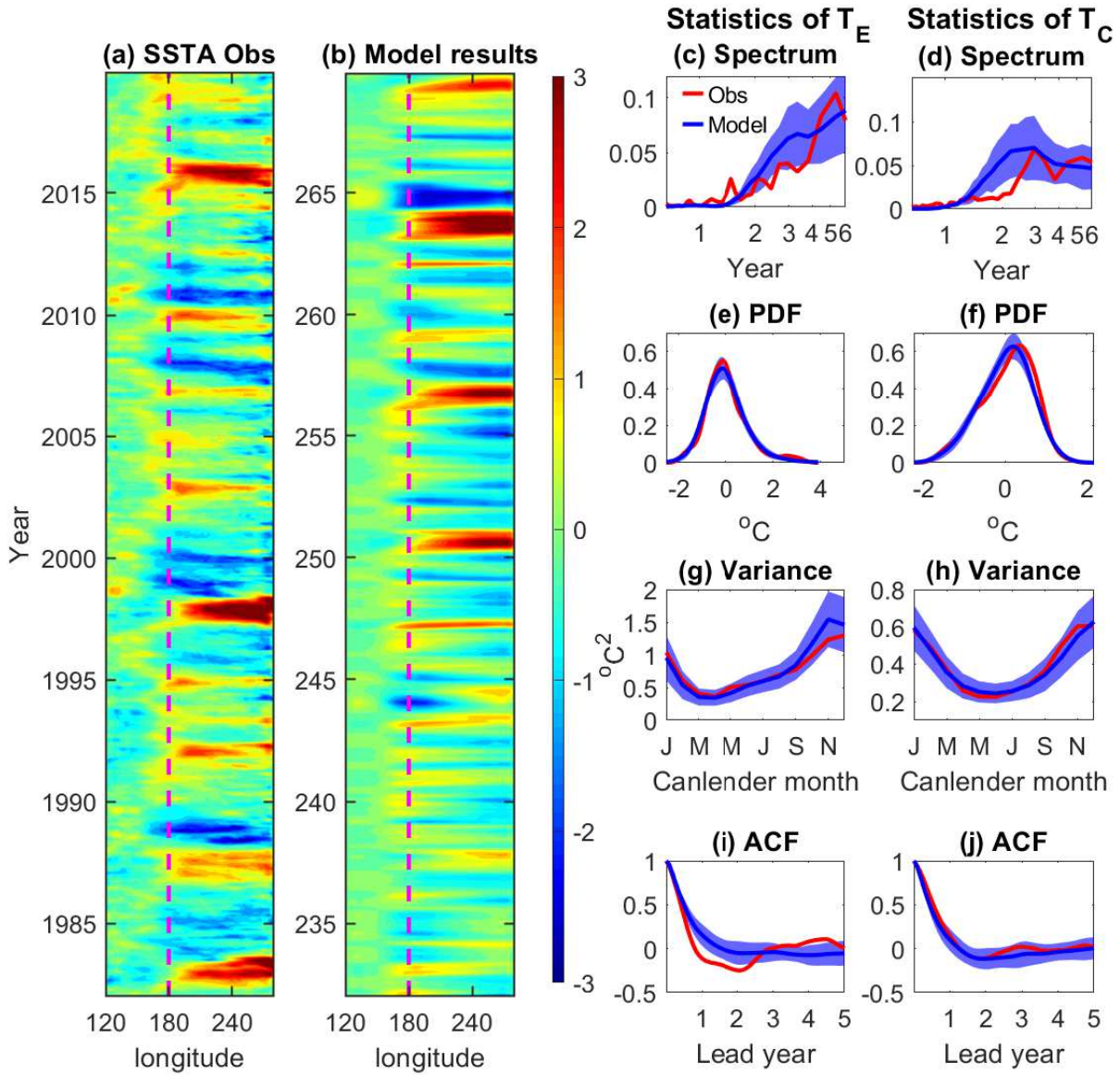}}
\caption{Comparison of the spatiotemporal patterns and the statistics between the observations and the coupled multiscale stochastic model. Panel (a) is the SST anomalies along the equatorial Pacific during 1982-2019. Panel (b) is same as (a) but is the random selected period during model year 232-269. Panels (c) and (d) are power spectrums of Ni\~no3 and Ni\~no4 SST, respectively. Panels (e) and (f) are PDFs. Panels (g) and (h) are the monthly variance (i.e., the seasonal cycle). Panels (i) and (j) are the ACF. In each panel, red and blue curves are for the observation and model, respectively. For the model, the total 2000-year long simulation is divided into 54 non-overlapping segments, each of which has a 37-year period as the observation. Then the average (blue line) and its one standard deviation intervals (shading) are illustrated.} \label{statistics}
\end{figure}

As was studied in \cite{chen2022multiscale}, this model can reproduce many desirable dynamical and statistical features of the observed ENSO complexity. Figure \ref{statistics} illustrates a comparison between the model simulations and the observations. Panels (a)--(b) include the Hovmoller diagrams of the SST, which are reconstructed by a bivariate linear regression, where the regression coefficients at each longitude is provided by computing the correlations of its SST value with the time series of $T_E$ and $T_C$, respectively.
These two panels reveal that the model can reproduce different types of the ENSO events with distinct warming or cooling centers, amplitudes and durations, as are observed in nature. Panels (c)--(j) compare the model statistics with the observations in both the CP and the EP regions. The skillful recovery of the non-Gaussian climatological PDFs and the ACFs indicates that the model satisfies both the necessary conditions in the information theory based framework, namely the model fidelity and the model memory. In addition, the observed power spectra and the seasonal phase-locking features are captured by the model. 
Note that the ACF of CP decays slower than that of EP in observations, indicating a longer memory of the CP El Ni\~nos. The model can accurately reproduced the observed ACF of the CP, which is a first evidence implying the skill of the model in potentially reaching a longer predictability of the CP events. Generally, this model is statistically accurate and involves the basic large-scale dynamical features of the ENSO complexity, which provides a reasonable justification for applying it to the study of the predictability within the information theory based framework. The only statistical inconsistency with the observation is that the ACF of the $T_E$ in the model decays slightly faster than the observations, which underestimates the observational biannual tendency that is probably associated with the effective El Ni\~no to La Ni\~na transitions.

\begin{figure}[h]
\centerline{\includegraphics[width=\textwidth]{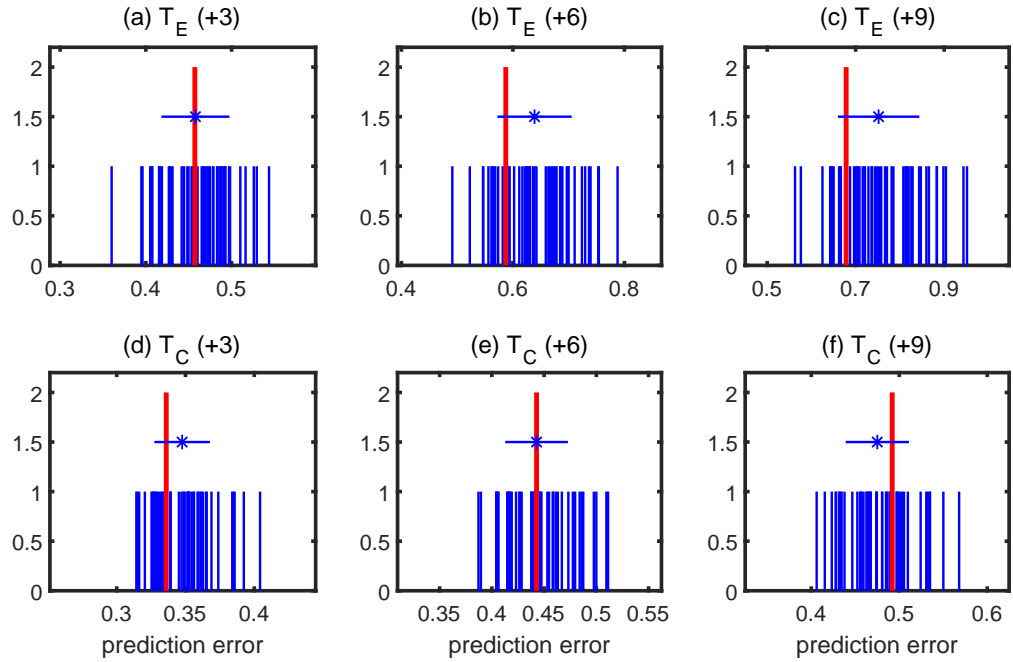}}
\caption{Prediction error measured for the observation (1982–2019) and 54 non-overlapping segments of the model years with the same length. For each segment and the observation, 100 members are used to perform the ensemble prediction for every month with a lead time from 1 to 12 months . Then, the ensemble mean of the predictions is used to calculate the RMSE with the target. The number in the title of each subplot indicates the lead time in months. The value for the observation is plotted with a longer bar. The mean and standard deviation of the statistics obtained from the segments are represented by the horizontal error bars. The y axis has no physical meaning but clearly separates the observation and the segments.} \label{surrogates}
\end{figure}

Before that, it is necessary to conduct a simple test of the model’s representation of the observational forecast errors. For this, a 2000-year long model simulation is generated and it is divided into 54 non-overlapping segments, each of which has the same length of time as the observations, i.e., 1982–2019. In each segment, we forecast $T_E$ and $T_C$ based on 100 ensembles. For observations, we use the observed $u$, $h_W$, $T_C$, $T_E$, $\tau$ and $I$ as the initial values to forecast the following ENSO evolution in the observations. While for different model segments, the corresponding exact values from the simulation are utilized as the initialization to forecast the ENSO evolution in the corresponding segment. Figure \ref{surrogates} shows the forecast errors for each segment and the observation, i.e., the root mean square error (RMSE) between the forecast ensemble mean and the true fields. It is seen that there is no significant difference in the forecasting ability of ENSO between the observation and model for both $T_E$ and $T_C$. That is, the longer bars corresponding to the observations are all covered by the mean and one standard deviation of the statistics obtained from the model segments. This suggests that the model is also capable of producing very realistic ENSO properties from a forecasting perspective, which further strengthens the justification of utilizing the model in studying the predictability of the ENSO complexity.

\section{The Overall ENSO Predictability}\label{Sec:Overall}
\subsection{The overall predictability, its Gaussian approximation, and the signal-dispersion decomposition}
\begin{figure}[h]
\centerline{\includegraphics[width=1.2\textwidth]{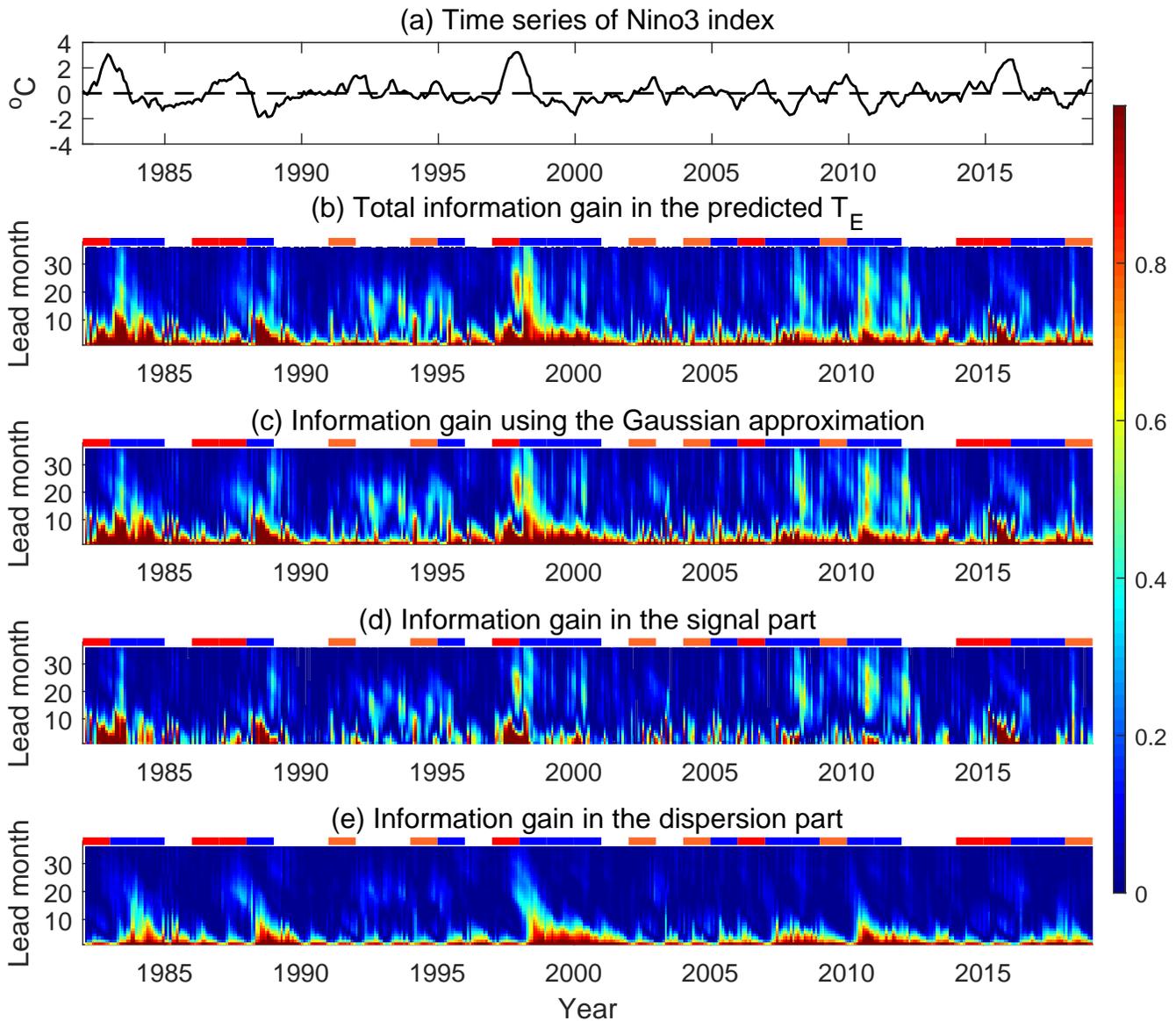}}
\caption{The information gain of the $T_E$ as a function of the starting date (x-axis) and the lead (y-axis). Panel (a) shows the time series of the observational Ni\~no3 ($T_E$) index. Panel (b) shows the total information gain based on the non-Gaussian PDFs. Panel (c) is the information gain computed based on the Gaussian approximations. Panels (d) and (e) show the signal and dispersion components of the information gain, respectively. The solid red line, orange line and blue line at the top of each panel represents EP El Ni\~no, CP El Ni\~no and La Ni\~na years, respectively.} \label{RE_standard_TE}
\end{figure}

\begin{figure}[h]
\centerline{\includegraphics[width=1.2\textwidth]{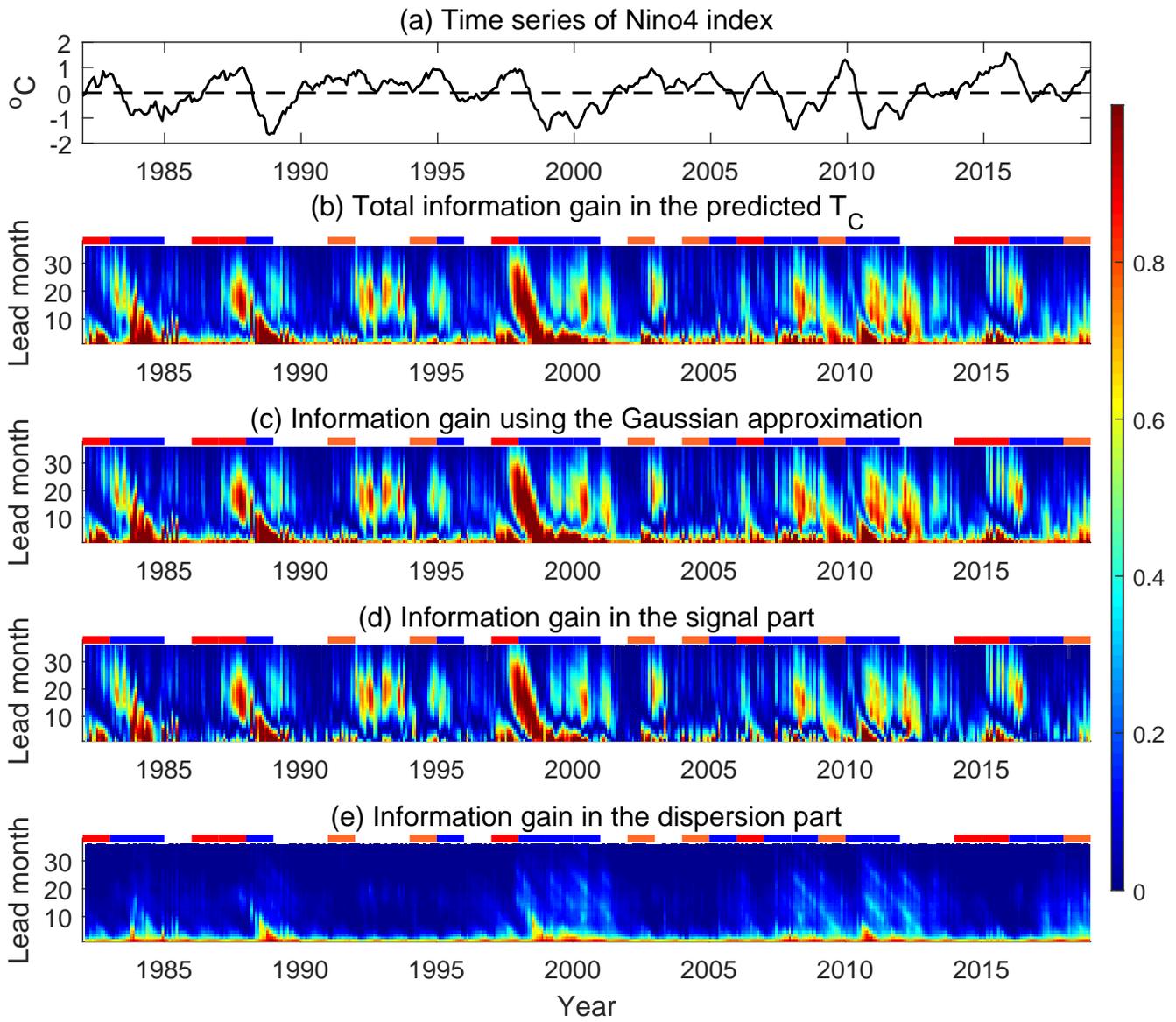}}
\caption{Similar to Figure \ref{RE_standard_TE} but for $T_C$.} \label{RE_standard_TC}
\end{figure}

Figures \ref{RE_standard_TE} and \ref{RE_standard_TC} show the information gain of $T_E$ and $T_C$, respectively, as a function of the starting date (x-axis) and the lead time (y-axis). An ensemble forecast by running the multiscale stochastic model \eqref{Model_Stochastic} forward is utilized to compute the time evolution of the PDF, where the observation data of all the six variables are directly adopted as the initial value. Since the model is a low-dimensional system, a large number of the ensembles containing $2000$ members is utilized, which ensures that the sampling error due to the insufficient ensemble number is negligible. In Figures \ref{RE_standard_TE} and \ref{RE_standard_TC}, the solid red, orange and blue lines at the top of each panel mark the years with EP El Ni\~no, CP El Ni\~no and La Ni\~na, respectively. To clarify the definition of ``an(a) EP El Ni\~no/CP El Ni\~no/La Ni\~na year", an entire calendar year is treated provided that such an event is observed in the year even if it lasts only for a couple of months (especially at the end of the year corresponding to the boreal winter). The following conclusions can be drawn from these two figures.

First, the time evolution of the information gain, namely the predictability, varies significantly as a function of the starting date. In particular, the predictability limit of the events starting from a La Ni\~na year is longer than those starting from both the CP and the EP El Ni\~no years. Yet, different individual events display distinct predictability limits. Such a finding has a remarkable significance for the study of the ENSO predictability. It implies that it is too crude to employ a single value of the predictability limit for the entire ENSO, which is however what the time series-based measurements can often provide by averaging over the entire testing period. Instead, it is anticipated that the ENSO predictability needs to be quantified for each individual event, which will be analyzed in detail in Section \ref{Sec:Complexity}. In addition, during the normal years, it is always not very predictable, which emphasizes the importance of the initial conditions related signals \citep{planton2021evaluating}. The seasonal dependency of the predictability will be discussed in Section \ref{Sec:Overall}b.

Second, comparing Panel (a) in the two figures, it can be seen that $T_C$ overall has stronger predictability than $T_E$. This is seemingly counterintuitive as the strength of the CP ENSO events is often weaker than that of the EP ones. However, special cautions need to be paid here: EP and CP ENSO events do not solely correspond to $T_E$ and $T_C$ indices. Both events have composite variations of them. The reason for the stronger predictability of $T_C$ is that its time series has a stronger temporal correlation than that of $T_E$ (see Figure \ref{statistics}). In fact, it has been shown that the skillful prediction (above the threshold of pattern correlation $= 0.5$) remains at a longer lead time in predicting $T_C$ than $T_E$ \citep{ren2019statistical,wang2020improving}.

Third, comparing Panels (a) and (b) in both figures, the Gaussian approximation seems to be an efficient and accurate simplification in computing the information gain, which is consistent with the finding in \citet{tang2005reliability}. Such a conclusion has a profound influence on applying the information theory to studying the ENSO predictability based on more sophisticated models, in which only a small number of the ensembles is affordable. A small ensemble size is often insufficient to recover the entire non-Gaussian PDF especially for high dimensional systems but it may provide a reasonably accurate Gaussian approximation one \citep{gardiner2009stochastic}. Yet, it is important to highlight that the skillfulness of the Gaussian approximation only implies that the information gain due to the higher order statistics behaves in a similar way as that from the mean and the variance. It does not suggest a linear model with Gaussian statistics is sufficient for the study of the predictability here. In fact, the information difference between the non-Gaussian climatological PDF (which reflects the asymmetric features of ENSO) and the optimal Gaussian approximation is quite significant, which means a linear model with Gaussian statistics will violate the model fidelity to a large extent. Nevertheless, the high similarity between the total information gain and its Gaussian fit allows us to simply consider the latter in the remaining of the paper, which also facilitates the discussions with its signal-dispersion decomposition.

Finally, from the signal-dispersion decomposition of the information gain, it can be concluded that the signal part dominates the total information gain in $T_C$ while both the signal and dispersion parts have remarkable contributions to $T_E$. This may suggest that the ensemble spread growth strongly depends on the initial conditions related signals, which is consistent with the finding in \citet{planton2021evaluating}. Note that the latter is different from the qualitative conclusion in \cite{tang2005reliability}, possibly because the model fidelity was not fully taken into account in that earlier work. The significant contribution from the dispersion part also indicates the importance of considering the entire forecast ensembles in characterizing the predictability instead of only focusing on the ensemble mean that merely impacts the signal part of the total information gain.

\subsection{The predictability as a function of the starting month}

\begin{figure}[h]
\centerline{\includegraphics[width=\textwidth]{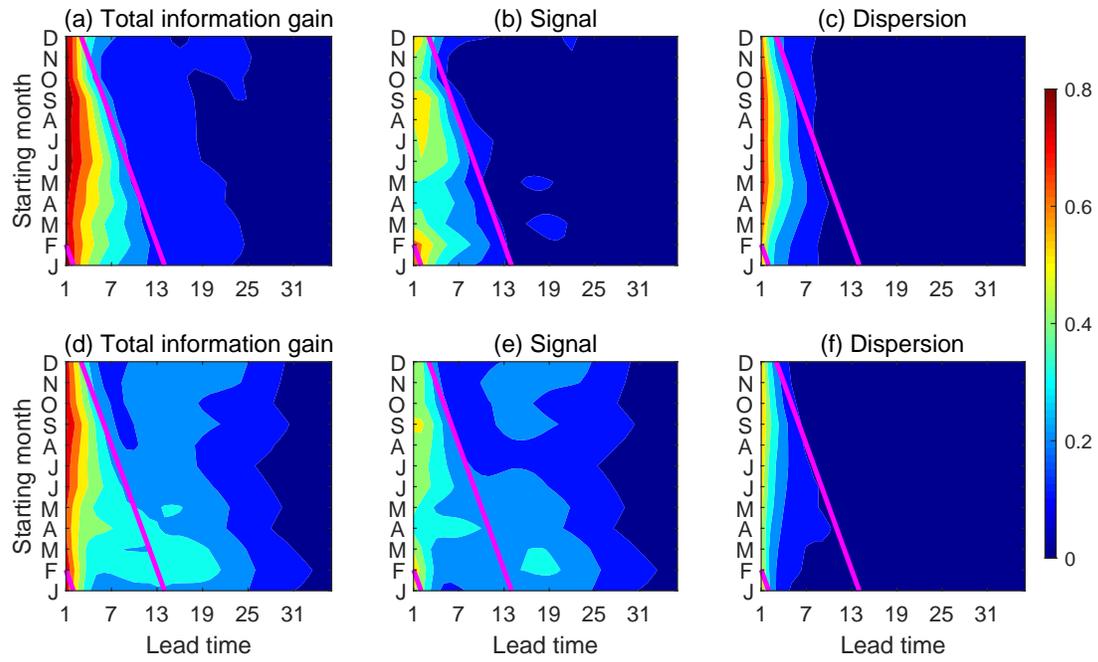}}
\caption{The information gain as a function of the starting month. Panels (a)--(c) are for $T_E$ and panels (d)--(f) are for $T_C$. The pink oblique lines in each panel are used to delineate the start of the boreal spring.} \label{TE_TC_composite}
\end{figure}

Figure \ref{TE_TC_composite} shows the predictability performance of $T_E$ and $T_C$ as a function of the starting month. The results confirm that $T_C$ has overall a stronger predictability than $T_E$ and that its main contributor is the signal part while the dispersion is a non-negligible component for predicting $T_E$. One particularly interesting finding from this figure is related to the well-known spring prediction barrier problem in many practical model forecasts and the persistence, which says the ENSO forecast skill drops significantly when straddling boreal spring. Yet, as for the information theory based predictability from the forecast statistics, the forecast around spring does not demonstrate any significant barrier. On the contrary, the overall predictability is even longer for the events starting from the spring season, especially the SST in the CP region. This indicates that if accurate and proper information can be obtained around spring, there might be a potential to make meaningful ENSO forecasts at long lead times, beyond the information provided by the climatology. Such a conclusion is also consistent with the results in a recent study \citep{fang2021effect}. It also highlights the difference between the time series-based prediction skill and the statistical based predictability. One interesting task is to apply other models in the ensemble forecast and information theory framework to explore if the spring barrier also disappears in the predictability.

\begin{figure}[h]
\centerline{\includegraphics[width=0.8\textwidth]{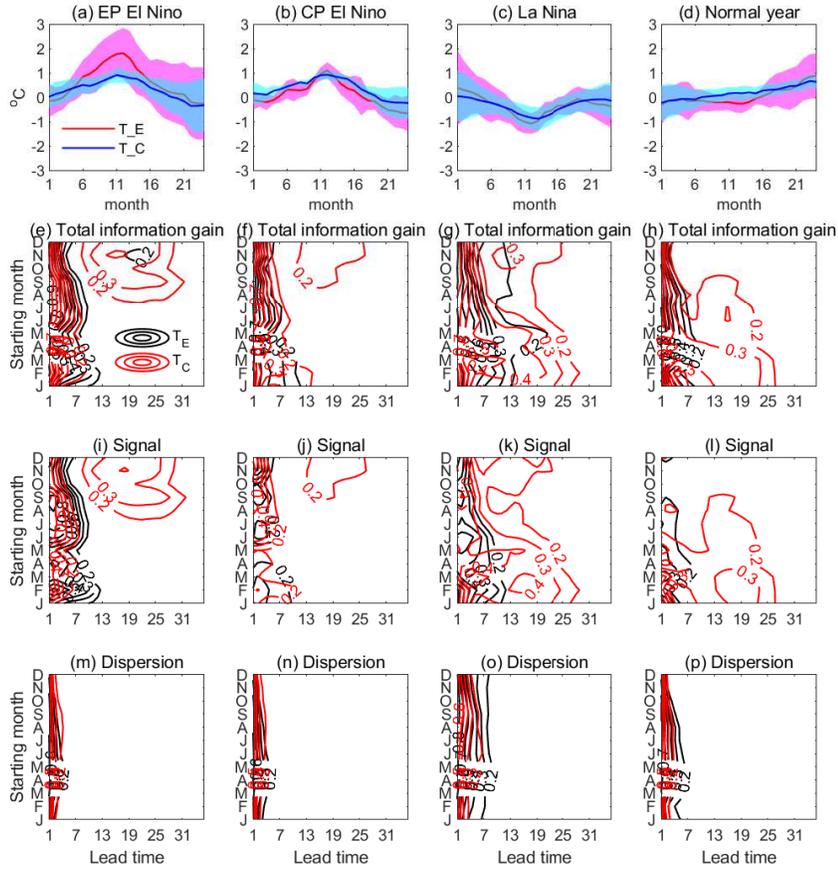}}
\caption{The information gain in predicting $T_E$ and $T_C$ as a function of the starting month for EP El Ni\~no, CP El Ni\~no, La Ni\~na and normal years. The top row shows the composite evolutions (solid lines) and their one standard deviation (shadings) of each event in the observations. The red and blue ones are for $T_E$ and $T_C$, respectively. The second, third and bottom rows illustrate the total information gain, the signal part and the dispersion part, respectively. In each panel, the black and red lines are the information gain in predicting $T_E$ and $T_C$, respectively.} \label{ENSO_composite}
\end{figure}

To further understand the difference in the ENSO predictability with different initial conditions, the four columns of Figure \ref{ENSO_composite} show the information gain in predicting $T_E$ and $T_C$ during the EP El Ni\~no, CP El Ni\~no, La Ni\~na and normal years, respectively. The additional messages provided by Figure \ref{ENSO_composite} are as follows.
First, a stronger predictability is observed if the forecast starts from a La Ni\~na year than a year with either type of the El Ni\~no, especially when the starting date ranges from January to June. The contribution from these La Ni\~na years accounts for the overall stronger predictability in the spring season, as was shown in Figure \ref{TE_TC_composite}, since the results associated with the El Ni\~no events are the opposite. The stronger predictability starting from a La Ni\~na year is probably related to the more effective discharge process of El Ni\~no at its mature stage \citep{planton2018western}. Quantitatively, El Ni\~no has a predictability of 6–10 months in general, while La Ni\~na has a much longer one, which can often exceed two years. Such a finding suggests that there remains a potentially large gap between our current level of ENSO forecast and its potential predictability, which has a room for the further improvement.
Second, it is noticed that the information gain of $T_C$ (red curves) starting from the second half of the year for both the EP and CP El Ni\~no years has a clear reemergence. This is mainly related with the effective El Ni\~no to La Ni\~na transitions \citep{planton2021evaluating}, which can be clearly seen from their composited evolutions (Figures \ref{TE_TC_composite}a--b).
Third, the EP El Ni\~no events show a stronger reemergence at the longer lead time for the forecast that begins in the fall or winter than the CP events. In addition, the EP El Ni\~no generally has a slightly stronger predictability than the CP one. This is consistent with our intuition since the former usually has stronger amplitudes, or in other words, signals. It is also not contradictory with the conclusion drawn from Figures \ref{RE_standard_TE}--\ref{RE_standard_TC} that $T_C$ overall has stronger predictability than $T_E$. Combing with these messages, we can conclude that the ENSO predictability is tightly related with the initial conditions related signals, which is consistent with the finding by \citet{planton2021evaluating} that the ENSO predictability should be discussed as a function of initial conditions, rather than as a function of the future evolution of the system since which has a random part.

\section{Predictability of Different ENSO Events Consisting of the ENSO Complexity}\label{Sec:Complexity}
\begin{figure}[h]
\centerline{\includegraphics[width=0.7\textwidth]{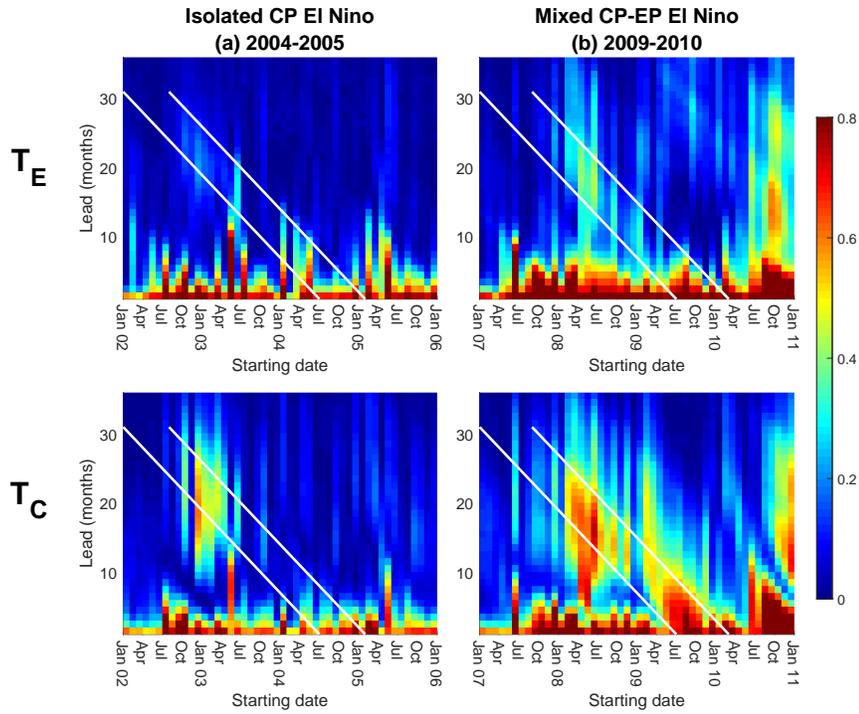}}
\caption{The total information gain in predicting $T_E$ and $T_C$ for an isolated CP Ni\~no event (2004-2005)  and a mixed CP-EP event (2009-2010). The two white solid lines provide the time window within which the corresponding event is active.} \label{complexity_CPEN}
\end{figure}

Figures \ref{complexity_EPEN}, \ref{complexity_CPEN} and \ref{complexity_LaNina} show the information gain in predicting 7 different ENSO events that consist of the ENSO complexity: 1) a moderate EP El Ni\~no (1986-1988), 2) a super El Ni\~no (1997-1998), 3) a delayed super El Ni\~no (2014-2016), 4) an isolated CP El Ni\~no (2004-2005), 5) a mixed CP-EP El Ni\~no (2009-2010), 6) a single-year La Ni\~na (1988-1989) and 7) a multi-year La Ni\~na (1999-2000). See Section \ref{Sec:Data_Model}\ref{Subsec:Def_Events} for the definitions of these events.\medskip

\subsection{Moderate EP El Ni\~no, super El Ni\~no and delayed super El Ni\~no}

\begin{figure}[h]
\centerline{\includegraphics[width=\textwidth]{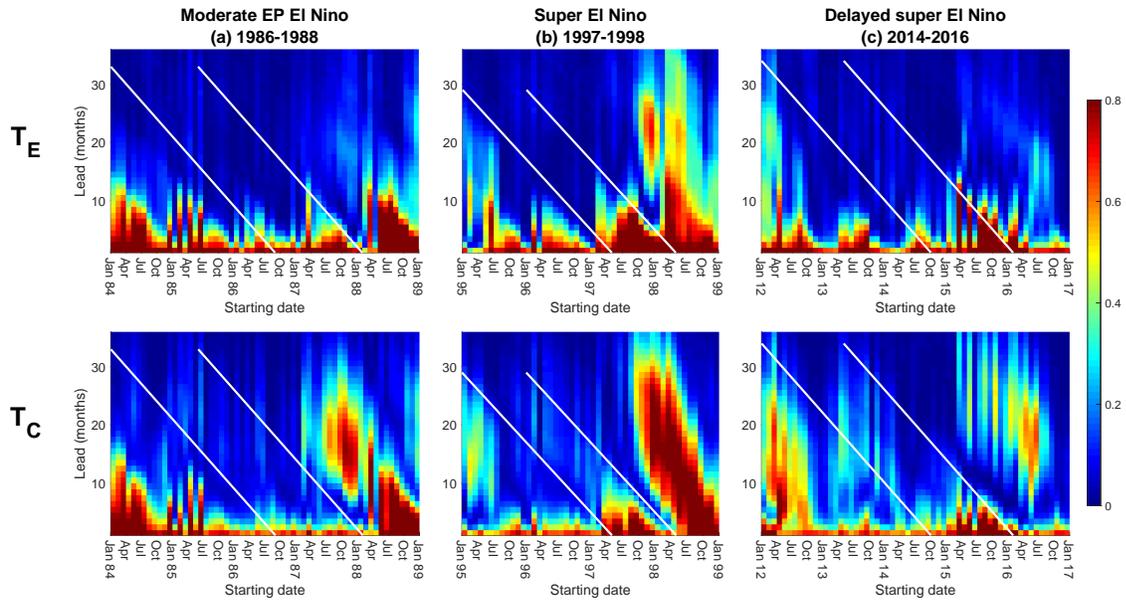}}
\caption{The total information gain in predicting $T_E$ and $T_C$ for a moderate EP El Ni\~no event (1986-1988), a super El Ni\~no event (1997-1998) and the delayed super El Ni\~no event (2014-2016). The two white solid lines provide the time window within which the corresponding event is active. } \label{complexity_EPEN}
\end{figure}

First, according to Figure \ref{complexity_EPEN}, the information gain along the first white solid line for all the three EP El Ni\~no events decays to zero very quickly, which indicates that it is very challenging to predict their onset phases. In fact, the westerly wind bursts are believed to be one of the major triggers of this type of events \citep{harrison1997westerly, tziperman2007quantifying, puy2016modulation}. However, they lie in the intraseasonal time scale and are inherently hard to be forecasted in the interannual time scale. This leads to the intrinsic difficulty in effectively predicting the onset of the EP El Ni\~no.

In addition to the common features in the predictability of all the three EP El Ni\~no events, there are also some differences among them. Column (a) of Figure \ref{complexity_EPEN} shows the predictability of the 1986-1988 moderate EP El Ni\~no event. When the starting date is at the growing phase (i.e. October 1986 to March 1987), the significant value of the information gain maintains only for about 6 months. When the starting date is after the event peak, i.e., July 1988, the predictability starts to become longer. This is because the initial value of the SST at these time instants is stronger than the climatological mean value, which provides additional important information that facilitates the prediction. One notable finding is that, if the starting time is between July 1988 to January 1989, then the information gain of $T_C$ within the first 10 months is close to zero. Nevertheless, the information gain has a significant increase afterwards and peaks at the lead time of about 15 to 20 months. Such a peak time undoubtedly corresponds to the subsequent La Ni\~na event, i.e., the predictability is related to the effective discharge process. The time span before the information gain reemergence corresponds to the phase change time from El Ni\~no to La Ni\~na and the time for the information passing from the EP region to the CP area.
Different from the moderate EP El Ni\~no event, Column (b) shows that starting from both the growing phase (April) and the mature phase (October) of the 1997-1998 super El Ni\~no, the information gain is significant for more than 10 months, though the gain is more pronounced for a short lead time in the latter case which is expected. This indicates the potentially stronger predictability of super El Ni\~no than the moderate events, even starting from the late spring season, which is also consistent with the current prediction skill using various models.

\begin{figure}[h]
\centerline{\includegraphics[width=\textwidth]{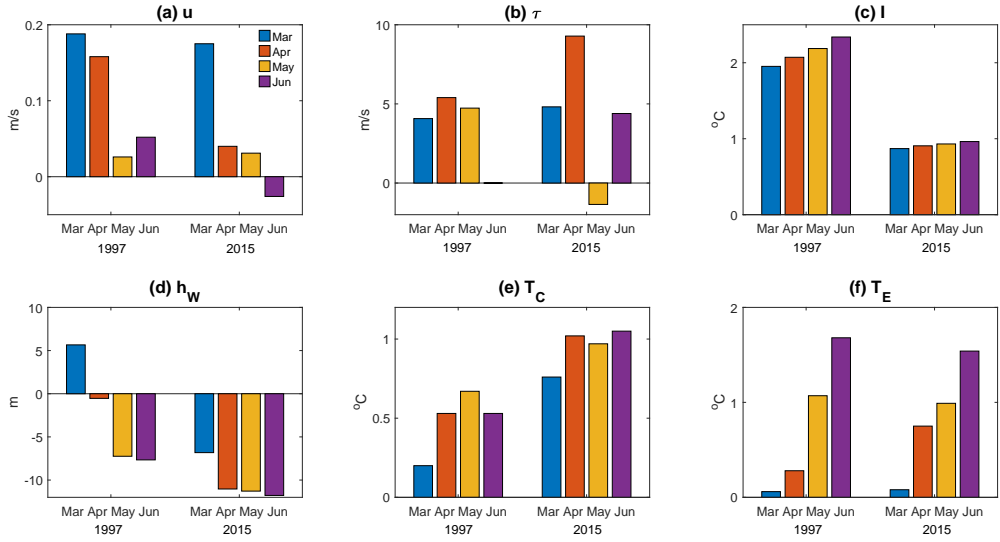}}
\caption{The initial values of different variables for the ensemble forecasts of super El Ni\~nos in 1997 and 2015. } \label{percentile_variables}
\end{figure}

\begin{figure}[h]
\centerline{\includegraphics[width=\textwidth]{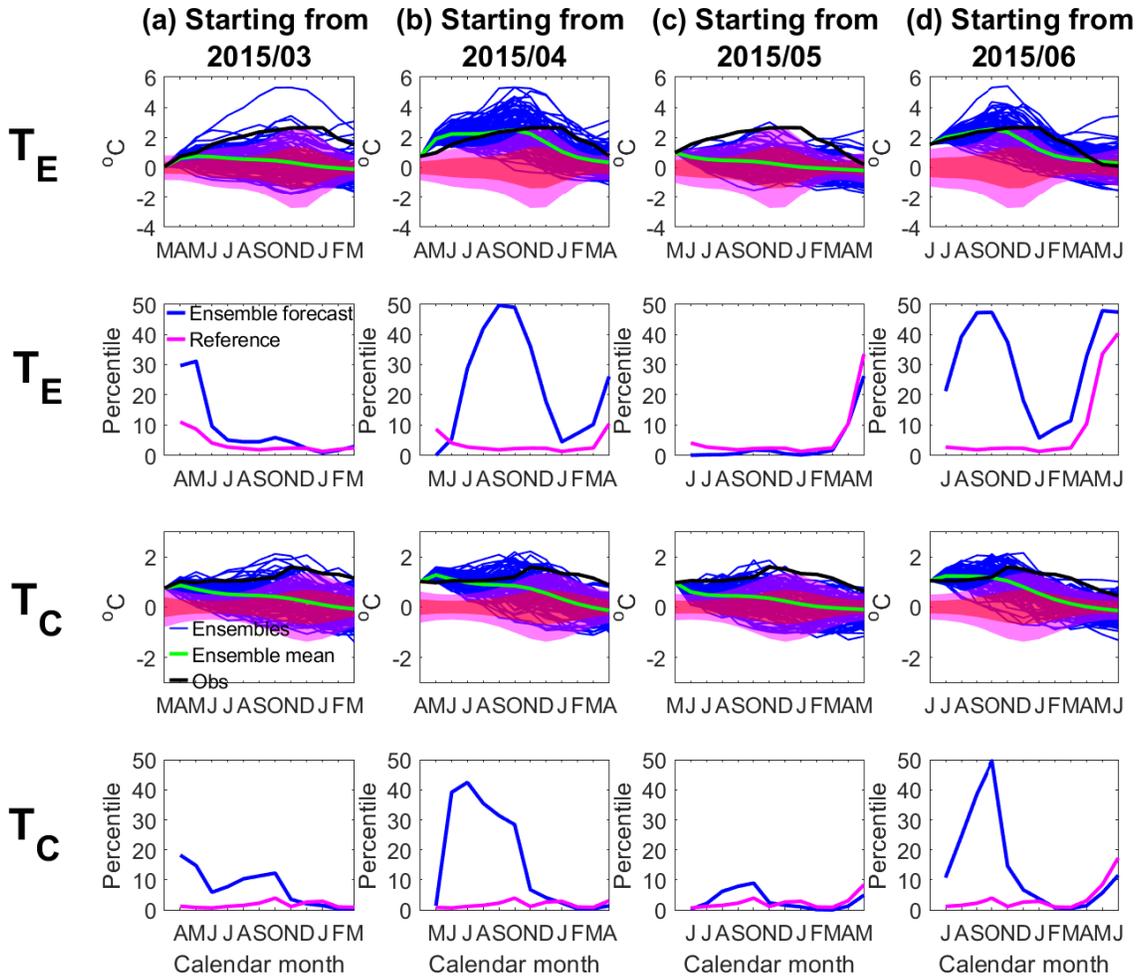}}
\caption{Prediction of the 2015-2016 super El Ni\~no. The first row shows the ensemble forecast of $T_E$, where the ensemble members are shown in blue curves and the ensemble mean is in green. It is compared with the true event, which is in black. The red and pink shading areas indicate the one and two standard deviations of the climatological PDF. The second row shows the percentile that the true event lies within the PDF of the ensemble forecast (blue) and that of the climatology (pink). The third and the fourth rows are similar but for $T_C$. The four columns show the ensemble forecasts starting from different months: (a) March, (b) April, (c) May, and (d) June. } \label{percentile_2015}
\end{figure}

\begin{figure}[h]
\centerline{\includegraphics[width=\textwidth]{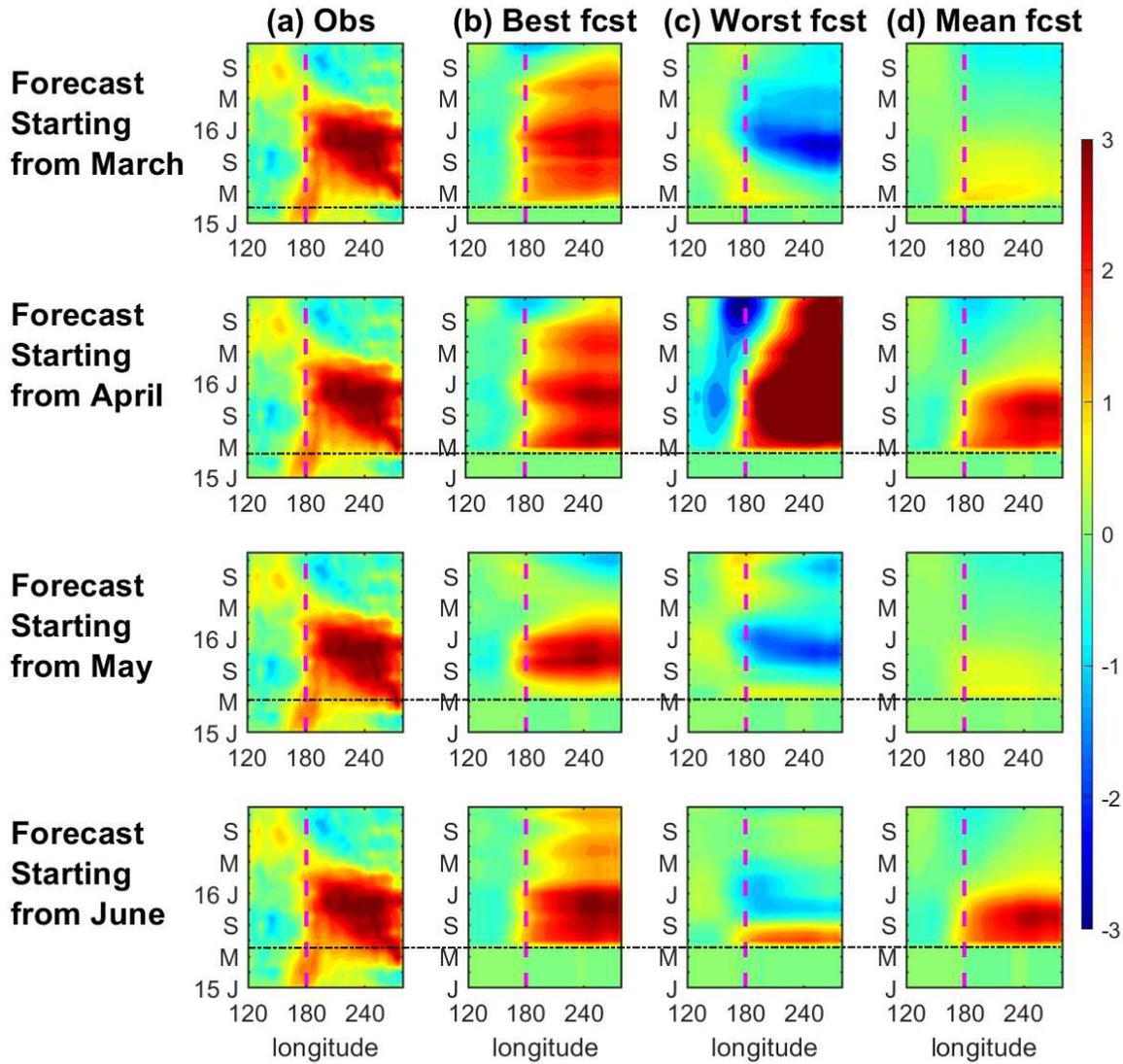}}
\caption{Hovmoller diagram of the ensemble forecast for the 2015-2016 super El Ni\~no. Column (a) is the true observed event. Column (b) and (c) show the best and the worst ensemble forecast member, respectively. Column (d) shows the ensemble mean forecast. Different rows show the forecast starting from different months, as marked by the black dashed-dot line. The values before the black dashed-dot line have been set to be zero. The pink dashed line is the dateline.} \label{percentile_2015_hov}
\end{figure}

Column (c) of Figure \ref{complexity_EPEN} shows the result of the 2014-2016 delayed super El Ni\~no. It can be seen that the predictability reaches a local peak if the starting date is around July 2014, which corresponds to the mature phase of the moderate El Ni\~no event in this three-year episode. Then regarding the subsequent super El Ni\~no in 2015, the information gain has a similar tendency as that in 1997-1998. The only difference is that the information gain of the 2015 event seems to be sensitive to the starting date. Specifically, the information gain remains significant if the starting date is April or June 2015 while the gain is consistently tiny regardless of the forecast lead time if the starting date is March, May or July. This is related to the rapid change of the wind bursts in 2015 \citep{chen2015strong,hu2016exceptionally,  thual2019statistical, xie2020unusual}. In fact, as is shown in Figure \ref{percentile_variables}, the westerly wind bursts suddenly disappear in May 2015 while other variables stay in the consistent states. As a consequence, there is no mechanism in lifting the ensemble members towards extreme values, which can be seen in Column (c) of Figure \ref{percentile_2015}. The Hovmoller diagram in Figure \ref{percentile_2015_hov} validates such a finding. It also shows that the reconstructed spatiotemporal pattern is poorly predicted starting from May 2015 in the ensemble mean forecast while the forecast is much more accurate if the starting date is either April or June. Note that the ensemble forecast PDF starting from March or May 2015 relaxes quickly towards the climatology while that starting from April or June is very distinguishable from the climatological PDF with the help of the strong wind bursts that also account for a large percentile of the ensemble members to forecast the 2015 super El Ni\~no event. These ensemble evolutions explain the sawtooth profile in Column (c) of Figure \ref{complexity_EPEN} for the predictability and links the forecast skill with the predictability.

Finally, it is noticeable that the predictability of $T_E$ extends further in time than that of $T_C$ for all the three types of the EP El Ni\~no events when the starting date is before or during the events. This seems to reach an opposite conclusion compared with the overall predictability shown in Figures \ref{RE_standard_TE}--\ref{RE_standard_TC}, which says $T_E$ has a weaker potential predictability than $T_C$. Yet, as was discussed in the previous section, the latter is due to the large contribution from the La Ni\~na events. Therefore, such a comparison again highlights the large discrepancy in the predictability of different types of the ENSO events and  indicates necessity of studying each type of the events.

\subsection{Isolated CP El Ni\~no and mixed CP-EP event}

Column (a) in Figure \ref{complexity_CPEN} shows the predictability of a single-year (or isolated) CP event in 2004-2005 while Column (b) displays that of a mixed CP-EP event in 2009-2010. When the forecast starts from anytime during the single-year CP event, the predictability remains quite weak. This is not surprising since the SST anomaly for most of the CP events is not as strong as the EP ones and therefore the additional information provided by the initial condition dissipated within a short time. In contrast, the predictability is more pronounced for the mixed CP-EP event, in which the strong EP SST contributes to the overall predictability and the information passes from the EP to the CP region.

In addition to these basic discoveries, there is one very interesting finding for both the events. That is, starting from about 20 months in advance (January 2003 and May 2008, respectively), the information gain can tell that the ensemble forecast distribution at the CP event peak is significantly different from the climatological PDF. Notably, this is not observed in the EP El Ni\~no or La Ni\~na events (Figures \ref{complexity_EPEN} and \ref{complexity_LaNina}) and it seems to be a unique feature of the CP El Ni\~no. Although a large information gain does not necessarily guarantee an accurate forecast, it does imply that there could be a potential to predict the occurrence of the CP events about two years in advance if suitable improvement is implemented in the current forecast systems.

\begin{figure}[h]
\centerline{\includegraphics[width=1.2\textwidth]{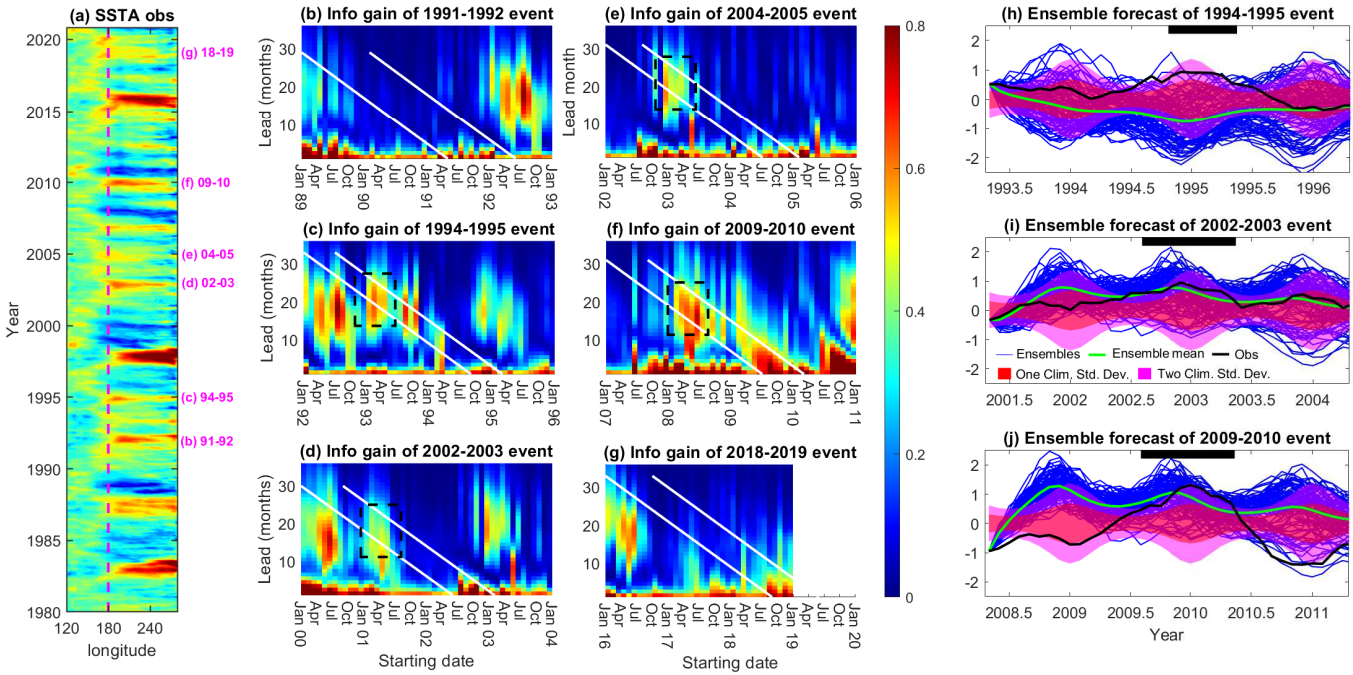}}
\caption{Information gain in predicting $T_C$ for different CP El Ni\~no events. Panel (a) shows the Hovmoller diagram of the observed SST anomaly. Six CP or CP-EP mixed El Ni\~no events are marked next to the Hovmoller diagram. Panels (b)--(g) show the information gain of $T_C$ for each of these events. The black dashed boxes inside some of these panels indicate the significant information gain at around 20 months lead time. Panels (h)--(j) show the ensemble forecasts for three of those six events, where the black bar at the top of each panel marks the time span of the corresponding CP El Ni\~no event. } \label{complexity_CPEN_all}
\end{figure}

To understand if the information gain remaining significant at the 20-month lead time is a universal characteristic for all CP events, Figure \ref{complexity_CPEN_all} shows the information gain in predicting $T_C$ for $6$ different CP El Ni\~no or CP-EP mixed events, which are marked next to the Hovmoller diagram in Panel (a). Among these $6$ events, shown in Panels (b)--(g), $4$ of them clearly demonstrate such a feature; they are the one in years 1994-1995, 2002-2003, 2004-2005 and 2009-2010. The CP event during 2018-2019 also reveals a tendency of the predictability at this long lead time, although the information gain is quite weak. On the other hand, the CP event during 1991-1992 does not display strong predictability. However, the warm SST center of the 1991-1992 event locates more towards the EP region compared with the other CP events, which might be the reason for the distinct behavior of the information gain of this event.

Panels (h)--(j) show the ensemble forecast for the 1994-1995, 2002-2003, and 2009-2010 events. Clearly, the ensembles at a lead time around 20 months is quite distinguishable from the climatology, which confirms a large information gain.
However, it is worthwhile to highlight again from these results that the predictability does not necessarily mean the skillful prediction. In fact, despite a large information gain, the ensemble mean time series is far from the observations during 1994-1995 event in Panel (h). Fortunately, the observed event is still captured by some ensemble members. Therefore, if a large information gain is obtained, then it is justified to conclude the forecast is distinguishable from the climatology while each plausible event from the forecast is assigned with a certain probability.  On the other hand, the ensemble evolutions in Panel (i) show a skillful forecast for the 2002-2003 event, converts the predictability to the actual prediction skill, while those in Panel (j) also accurately predict the occurrence of the 2009-2010 CP El Ni\~no.

\subsection{Single-year and multi-year La Ni\~nas}

\begin{figure}[h]
\centerline{\includegraphics[width=0.8\textwidth]{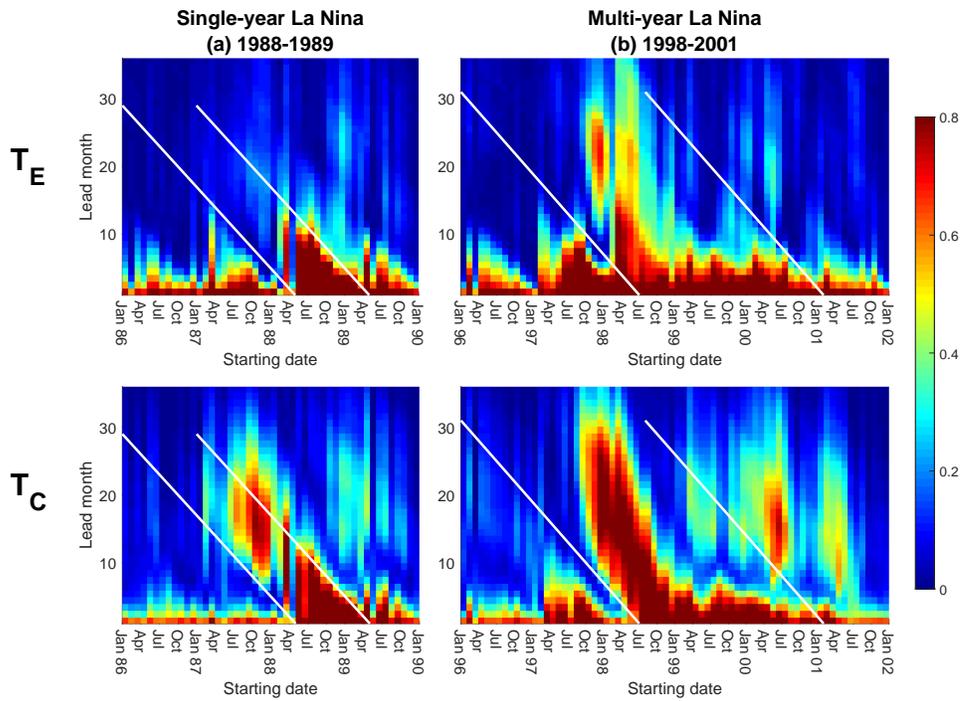}}
\caption{The total information gain in predicting $T_E$ and $T_C$ for a single year La Ni\~na event (1988-1989)  and a multi-year La Ni\~na event (1998-2001). The two white solid lines provide the time window within which the corresponding event is active.} \label{complexity_LaNina}
\end{figure}

The predictability of a single-year (1988-1989) and a multi-year La Ni\~na event (1998-2001) is shown in Figure \ref{complexity_LaNina}. The results here confirm the conclusion discussed in the previous section that the La Ni\~na events usually have stronger predictability since they are the discharge phase of the ENSO cycle. One interesting finding in the 1988-1989 one is that the bound of the significant value of the information gain is consistent with the La Ni\~na's demise (the white line). This implies the forecast ensembles reaching the climatological PDF is in phase with the relaxation of the La Ni\~na towards the quiescent state, i.e., this La Ni\~na event follows exactly the discharge-recharge paradigm and is therefore predictable. On the other hand, the 1998-2001 multi-year La Ni\~na shows a strong predictability when the forecast begins from either the preceding El Ni\~no event, even with the spring starting time, or the onset of the negative SST phase (middle of year 1998). The information gain decays as the starting time becomes 1999 or 2000, at which the amplitude of the initial value becomes weak.

\begin{figure}[h]
\centerline{\includegraphics[width=\textwidth]{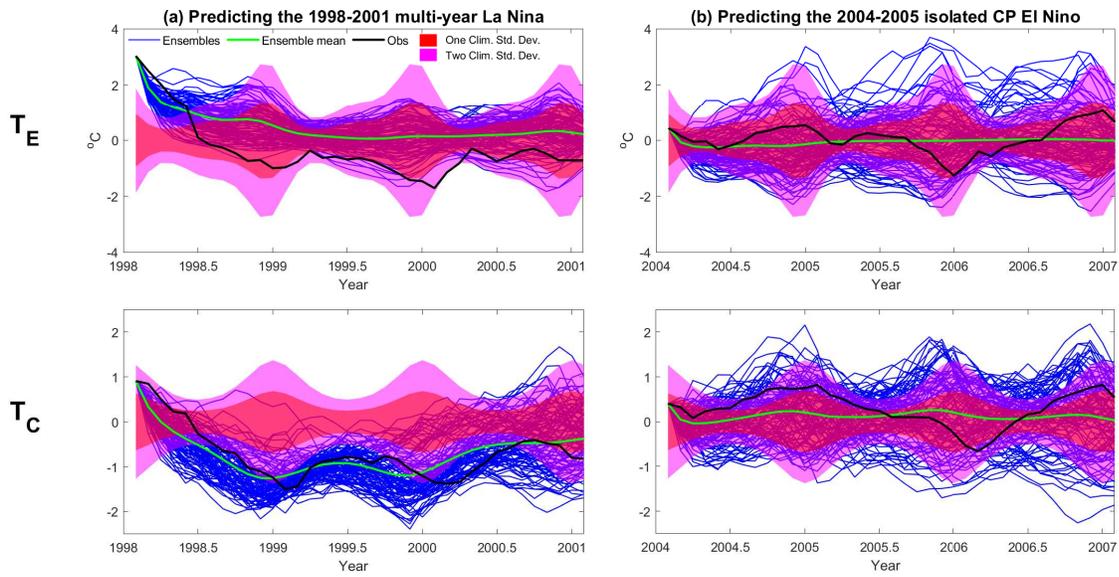}}
\caption{Ensemble forecast of the 1998-2001 multi-year La Ni\~na and the 2004-2005 isolated CP El Ni\~no. The blue curves show the 50 randomly selected ensemble members with the ensemble mean being green. The true observed event is shown in black color. The red and pink shading areas show the one and two standard deviations of the climatology distribution.  } \label{EnsembleFcst_98_04}
\end{figure}

Finally, Figure \ref{EnsembleFcst_98_04} includes some intuitive results. It shows the ensemble forecast of the 1998-2001 multi-year La Ni\~na and the 2004-2005 isolated CP El Ni\~no. As is shown in Column (a), starting from January 1998, the ensembles of $T_C$ for the 1998-2001 event evolve in a very different way from the climatological PDF for more than two years, which accounts for the large information gain as was shown in Panel (a) of Figure \ref{complexity_LaNina}. In fact, the ensemble mean here also tracks the truth in an accurate fashion. Thus, the strong predictability in such a situation indeed converts to the skillful forecast. On the other hand, as is shown in Column (b), the initial values of both $T_E$ and $T_C$ at January 2004 are close to zero and the ensembles spread quickly towards the climatology as well. In such a case, despite that the time series of the true event is consistently included within the ensemble spread, the forecast ensembles cannot effectively provide any useful additional information beyond the climatology.

\section{Sensitivity Analysis}\label{Sec:Sensitivity}

\subsection{ENSO predictability with different strengths of the decadal variability}
\begin{figure}[t]
\centerline{\includegraphics[width=\textwidth]{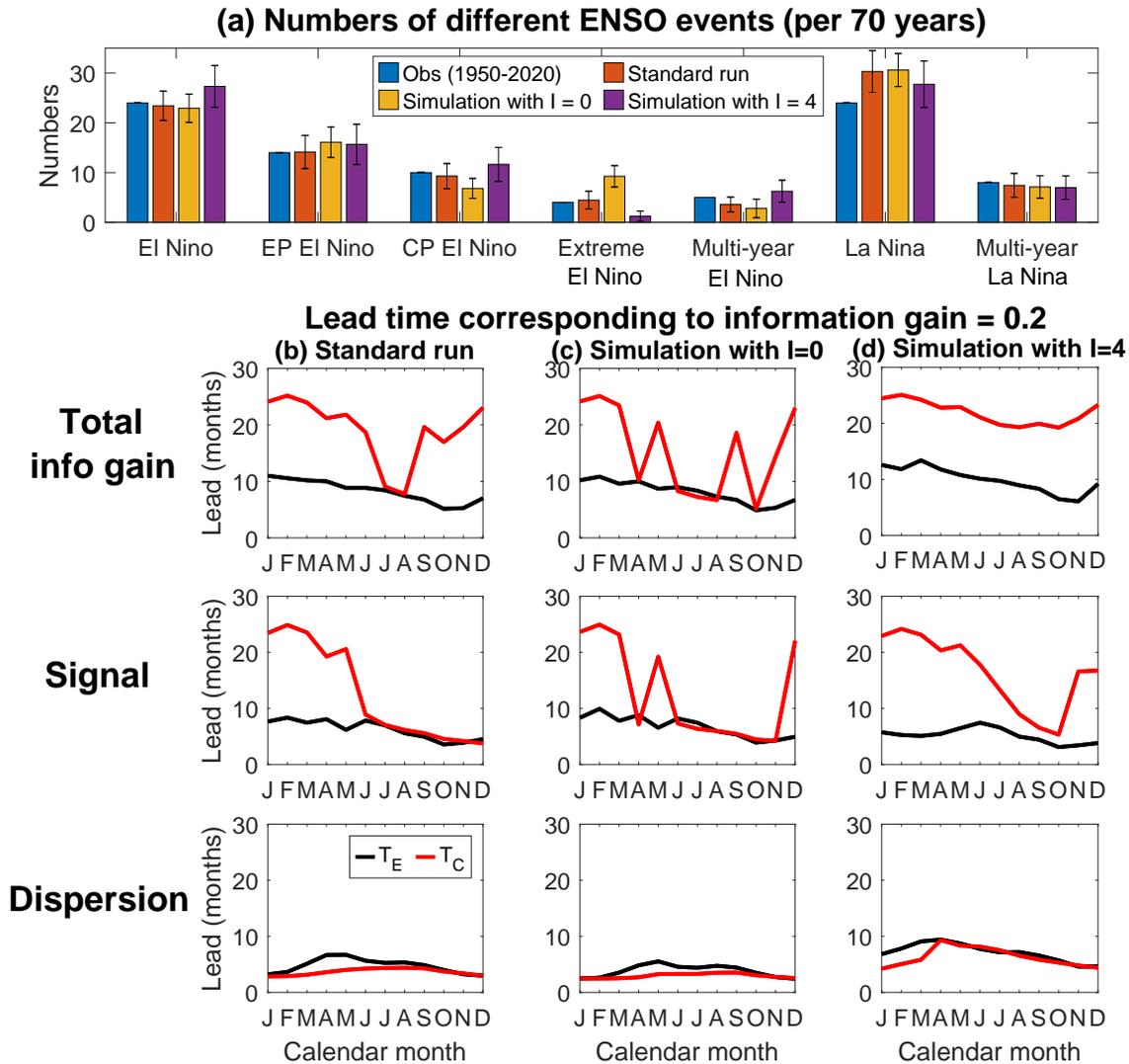}}
\caption{ENSO predictability with different strengths of the decadal variability using the twin experiments. Panel (a) shows the number of different ENSO events occurred per 70 years. The bar indicates the confidence interval based on 30 independent model simulations, each of which is 70 years long as the observations from 1950 to 2020. Panels (b)--(d) show the information gain in predicting $T_E$ (black) and $T_C$ (red) using the standard model run, the model with $I=0$ and the model with $I=4$. Here the threshold value $\mathcal{E}=0.2$ of the information gain in \eqref{Relative_Entropy_Rescaled} is used for comparison. } \label{RE_decadal_3cases}
\end{figure}

Recall, in the multiscale stochastic model \eqref{Model_Stochastic}, the variable $I$ represents the background dynamic Walker circulation and varies in the decadal time scale. One important practical issue is to understand the ENSO predictability in the climate change scenarios, which can be implemented by perturbing $I$ in the model.
The study in this subsection is based on perfect model twin experiments with different choices of the decadal variability $I$, which means the model is first used to generate synthetic time series as ``observations" and then the same model is used for quantifying the predictability. The reason to exploit this experiment is that the observational data of the possible future climate change scenario are not available. Since the model has been shown to be statistically accurate and the predictability of the model-generated time series has been validated to resemble that of the observations under the current climate (not shown here), the perfect model twin experiments are expected to at least provide some qualitatively useful conclusions.
In the following, three perfect model twin experiments are carried out:
\begin{itemize}
  \item [(a)] The current climate scenario. This is the standard run of the full model \eqref{Model_Stochastic} with the standard parameters that captures the current climate statistics, where $I$ is driven by the stochastic process \eqref{Model_Stochastic_I} and ranges from $0$ to $4$.
  \item [(b)] The climate scenario with the strengthening of the Walker circulation, i.e., the variable $I$ is set to be a constant $I=4$ in the model \eqref{Model_Stochastic} and the stochastic process \eqref{Model_Stochastic_I} is discarded.
  \item [(c)] The climate scenario with the weakening of the Walker circulation, i.e., the variable $I$ is set to be a constant $I=0$ in the model \eqref{Model_Stochastic} and the stochastic process \eqref{Model_Stochastic_I} is discarded.
\end{itemize}
The main results are illustrated in Figure \ref{RE_decadal_3cases}. Panel (a) shows the number of different ENSO events occurred per 70 years. To include the uncertainty quantification, the confidence intervals are also added to the bar plots. These confidence intervals are computed based on 30 independent model simulations, each of which is 70-year long as the observations from 1950 to 2020. As a further validation of the model, the occurrence frequency of each type of the ENSO events from the standard model run is compared with the observations under the current climate scenario. Despite a slight overestimation of the La Ni\~na events in the model, the occurrence frequencies of all the other types of the events match the observations very well, including the overall El Ni\~no events, the CP El Ni\~no events, the EP El Ni\~no events, the extreme El Ni\~no events and the multi-year events for both El Ni\~no and La Ni\~na. These comparisons provide reasonable justifications of utilizing the perfect model twin experiments to study the predictability in the climate change scenarios.

Panel (a) also shows that, with the strengthening of the background Walker circulation, the occurrence of the El Ni\~no events increases while that of the La Ni\~na decreases. Among different types of the El Ni\~no events, the occurrence of the CP El Ni\~no turns into more frequent while the extreme EP events become seldom to happen. On the other hand, although the total number of the La Ni\~na events decreases, the frequency of the multi-year La Ni\~na occurrence remains the same as the current climate. Therefore, according to the analysis of the predictability of the ENSO complexity in Section \ref{Sec:Complexity}, it is anticipated that the predictability of $T_C$ should increase as $I$ becomes large while that of $T_E$ remains at the same level as in the current climate. This conjecture is confirmed by Panels (b) and (d). In particular, the extended predictability of $T_C$ is mainly observed in summer and fall seasons. In addition, the predictability of $T_E$ is overall unchanged, although a slight increment of the dispersion part is found. Similarly, if the dynamic Walker circulation is constantly weakened (with $I=0$), then the predictability of $T_C$ decreases, especially in spring and fall (Panel (c)), while that of $T_E$ does not have obvious changes.

To summarize, in the climate change scenarios, the predictability of $T_C$ has the most significant response. When the background Walker circulation becomes stronger (weaker), the predictability of $T_C$ increases in summer and fall (decreases in spring and summer), while the predictability of $T_C$ in winter and that of $T_E$ throughout the year remains almost unchanged.

\subsection{Role of the multiscale components in affecting the ENSO prediction}

The focus of this subsection is on studying the influence of different model variables on the ENSO forecast, especially those that are not included in the classic recharge-discharge theory but play a crucial role in ENSO complexity. These variables include the intraseasonal zonal wind stress $\tau$, the decadal variable $I$ (proportional to the strength of the zonal advective feedback), and the zonal current $u$.

Such a study can also naturally fit into the information theory based framework with a slight rearrangement of the PDFs in computing the information distance,
\begin{equation}\label{Relative_Entropy_Model_Error}
  \mathcal{P}(p_{reduced}(t,u),p(t,u)) = \int p_{reduced}(t,u)\log\left(\frac{p_{reduced}(t,u)}{p(t,u)}\right)du.
\end{equation}
In \eqref{Relative_Entropy_Model_Error}, the reference solution $p(t,u)$ is the standard run of the ensemble forecast utilizing the full model \eqref{Model_Stochastic} while the PDF $p_{reduced}(t,u)$ comes from running a ``reduced model" in which one of the above mentioned variables is set to be zero. Then the information difference from \eqref{Relative_Entropy_Model_Error} (after rescaling using \eqref{Relative_Entropy_Rescaled}) is the information loss by ignoring the contribution from the specific variable.

\begin{figure}[h]
\centerline{\includegraphics[width=\textwidth]{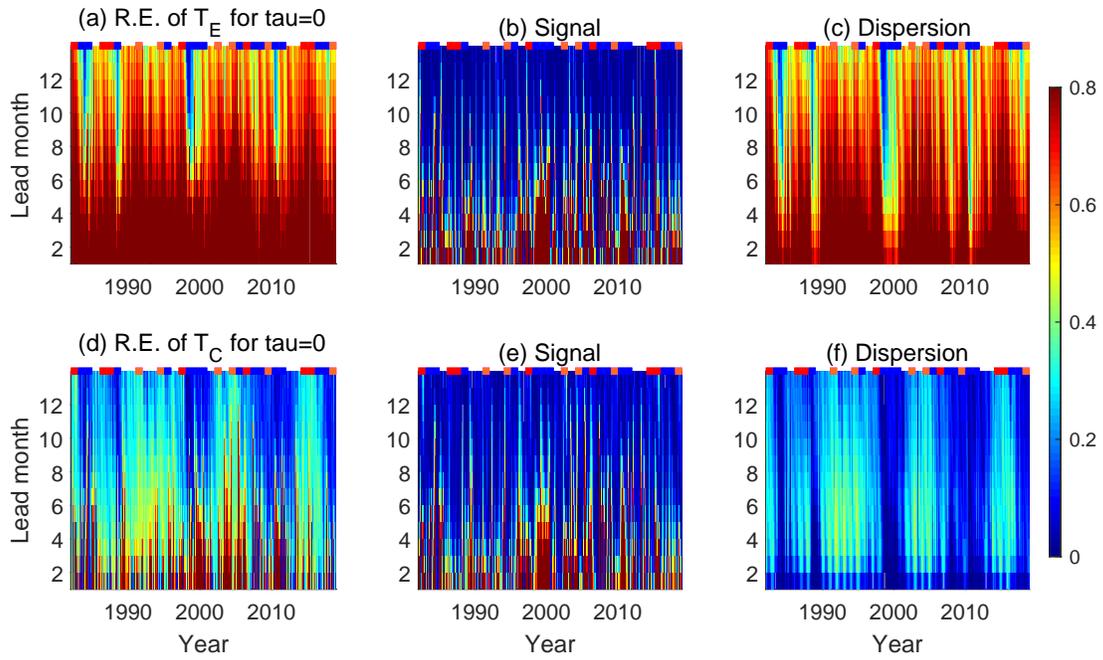}}
\caption{The information loss in predicting $T_E$ (Column (a)) and in predicting $T_C$ (Column (b)) due to the ignorance of the intraseasonal zonal wind stress $\tau$.  } \label{relative_importance_tau}
\end{figure}

It has been pointed out that the wind bursts are crucial to the ENSO development, especially to the SST in the EP region \citep{harrison1997westerly, tziperman2007quantifying, puy2016modulation}. This is confirmed by the information theory, where an enormous information loss is found by ignoring $\tau$ in the model ensemble forecast (Figure \ref{relative_importance_tau}). Notably, the composites of the loss of information in predicting $T_E$ and $T_C$ are very different. The loss of information is mainly on the dispersion part for predicting $T_E$, which means the ensemble spread is significantly underestimated. Therefore, although the EP SST is dominated by the interannual air–sea interactions, the intraseasonal information provides as a supply by the up-scale cascade, which means that a large portion of the variability in $T_E$ comes from $\tau$. In other words, $\tau$ triggers many events in the EP area, especially the extreme El Ni\~nos \citep{hu2016exceptionally}. On the other hand, although the role of $\tau$ on $T_C$ is not as strong as that on $T_E$, it still has a pronounced impact on the information gain in the CP area. Different from the prediction of $T_E$, the information loss in predicting $T_C$ by excluding $\tau$ is mainly reflected in the signal part. This implies the random wind bursts can influence the deterministic component of the dynamics of $T_C$ and lead to a mean bias in the forecast.

\begin{figure}[h]
\centerline{\includegraphics[width=\textwidth]{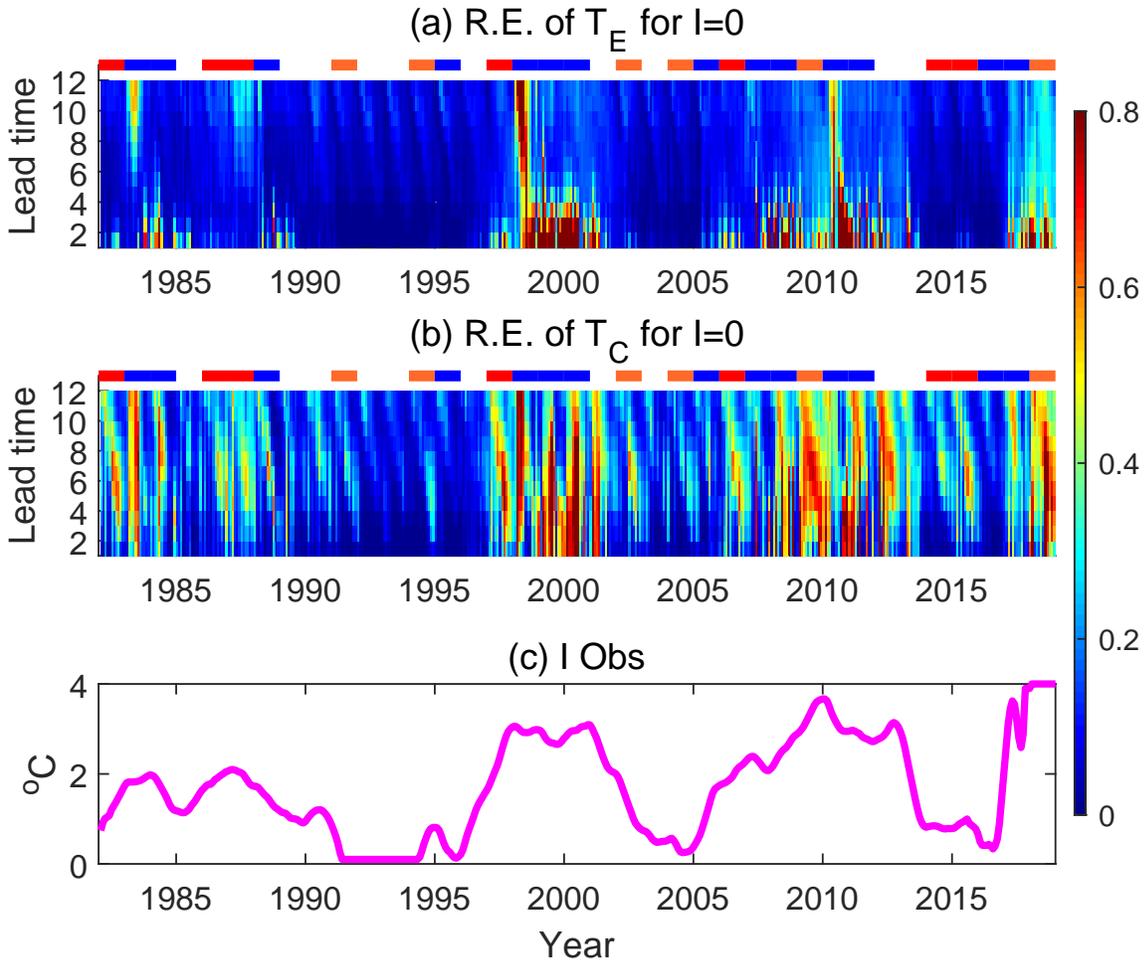}}
\caption{The information loss in predicting $T_E$ (Panel (a)) and in predicting $T_C$ (Panel (b)) due to the ignorance of the decadal variability $I$. For reference, the observed $I$ is shown in Panel (c).} \label{relative_importance_I}
\end{figure}

Next, Figure \ref{relative_importance_I} explains the impact of the decadal variability $I$ on the ENSO predictability. It is seen that $I$ modifies both $T_E$ and $T_C$ with a significant and synchronized decadal variation feature. In particular, strong influence of $I$ on the SST is found around 2000, 2010 and 2018, which are the years with La Ni\~na and CP El Ni\~no events. Therefore, although the ENSO events lie in the interannual time scale, the decadal variability is important for improving the predictability of ENSO in both CP and EP regions.

\begin{figure}[h]
\centerline{\includegraphics[width=\textwidth]{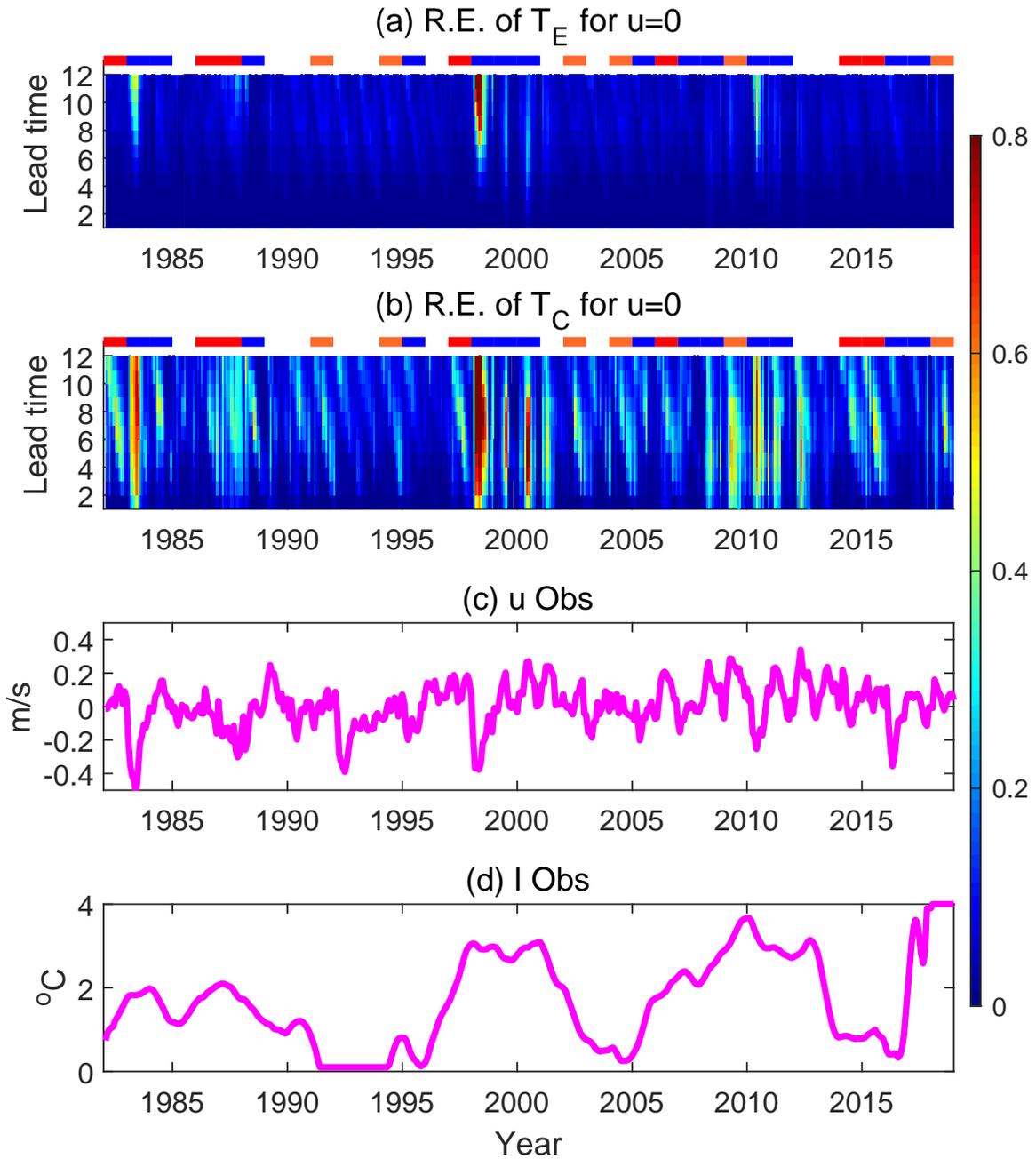}}
\caption{The information loss in predicting $T_E$ (Panel (a)) and in predicting $T_C$ (Panel (b)) due to the ignorance of the ocean current $u$. For reference, the observed $u$ together with the observed $I$ are shown in Panels (c) and (d).} \label{relative_importance_u}
\end{figure}

Finally, the loss of information by excluding the contribution from the zonal current $u$ in the CP is shown in Figure \ref{relative_importance_u}. The loss of information caused by ignoring $u$ in the forecast system is similar to that by disregarding the contribution from the decadal variability $I$, except that the influence of $u$ on the ensemble prediction is mainly found in the CP region. In fact, $u$ and $I$ are strongly correlated in driving the time evolution of $T_C$. Since $u$ is one of the main driving mechanisms of the CP El Ni\~no events, it has a direct contribution to the SST in CP area. On the other hand, the contribution of $u$ to the La Ni\~na is not as significant as that of $I$. Taking into account both Figures \ref{relative_importance_I} and \ref{relative_importance_u}, it can be concluded that, in the period of weak $I$ that leads to a minor contribution from the ocean zonal current, ENSO's development can be effectively described by only considering the vertical processes in the EP, e.g., the thermocline feedback. This is the main reason that in the 1980s and 1990s the simulations and forecasts of ENSO utilizing the traditional models with an emphasis on the thermocline feedback are very successful. However, in the period of strong $I$ (e.g., after 1999), ignoring the effect of the zonal current has a strong negative impact on the ENSO simulations and forecasts.

\section{Conclusions and Discussion}\label{Sec:Conclusions}
In this paper, the information theory based framework is applied to quantify the predictability of ENSO complexity, which includes different types of the ENSO events in both the EP and CP regions with diverse characteristics in spatial pattern, peak intensity, and temporal evolution. While most of the conventional models focus on describing certain key dynamical properties of ENSO, a recently developed multiscale stochastic model succeeds in capturing both the large-scale dynamics and many crucial statistical properties of the observed ENSO complexity, including the PDFs, the seasonal phase-locking, the power spectrums and the ACFs of the SST in both the EP and CP regions. These desirable features allow the model to be a unique statistically accurate dynamical system that facilitates the use of the information theory based framework to study the predictability of the ENSO complexity. Main conclusions are summarized as follows:
\begin{itemize}
  \item ~[\emph{Distinct predictability for different events}]. Different ENSO events possess very distinct predictability limits. Therefore, it is too crude to employ a single value of the predictability limit for the entire ENSO, as the time series based measurements often do for assessing the ENSO forecast skill.
  \item ~[\emph{Importance in both signal and dispersion}]. The CP SST $T_C$ overall has more predictability than the EP SST $T_E$. Although the signal part dominates the total information gain in $T_C$, both the signal and dispersion parts have remarkable contributions to $T_E$. The latter indicates the importance of considering the entire forecast ensembles in characterizing the predictability instead of focusing on only the ensemble mean.
  \item ~[\emph{No obvious spring barrier}]. The overall predictability starting from the spring season does not demonstrate any significant barrier. This indicates that if accurate and proper information can be obtained around spring, there might be a potential to make meaningful ENSO forecasts at long lead times, beyond the information provided by the climatology.
  \item ~[\emph{EP El Ni\~nos}]. The information theory based predictability indicates that  it is overall challenging to accurately predict the onset of the EP events, as the random wind bursts are one of its main triggering mechanisms. Both the predictability and the prediction skill may also differ significantly with a slight change of the starting date due to the rapid adjustment of the wind burst amplitude.
  \item ~[\emph{CP El Ni\~nos}]. There seems to be a universal tendency that, starting from about 20 months in advance of a CP El Ni\~no event, the time evolution of the information gain is always significant at the timing of the target CP event. Remarkably, the strong predictability indeed converts to the skillful forecast for predicting many CP events about 2 years in advance.
  \item ~[\emph{La Ni\~nas}]. Stronger predictability is found in the La Ni\~na events than both types of the El Ni\~nos, which is related to the more effective discharge process of ENSO at its mature stage.
  \item ~[\emph{Climate change scenario}]. In the climate change scenario with the strengthening of the background Walker circulation, the predictability of $T_C$ has a significant response with a notable  increase in summer and fall. The predictability of $T_E$ remains almost unchanged.
  \item ~[\emph{Role of different variables}]. The loss of information becomes significant for predicting EP El Ni\~no if the wind bursts are ignored and the main loss comes from the dispersion part, which is particularly detrimental to predicting the extreme events. The information loss in predicting $T_C$ by excluding the wind is also notable in the signal part, which may lead to a mean bias in forecasting $T_C$. In contrast, the ocean zonal advection mainly affects the predictability of the CP El Ni\~no. In addition, the decadal variability is important for improving the predictability of ENSO in both CP and EP regions.
  \item ~[\emph{Justification of the Gaussian approximation}]. The Gaussian approximation is shown to be efficient and accurate in computing the information gain. Such a justification facilitates the use of the information theory to studying the ENSO predictability based on more sophisticated models, such as the ICMs and CGCMs, for which only smaller ensemble sizes are affordable.
\end{itemize}

It is worthwhile to notice that the conventional time series-based measurements, such as the anomaly correlation coefficient and the RMSE of the ensemble mean time series, are simple and practically useful for assessing the ENSO forecast skill. Yet, they may not be the most suitable metric for quantifying the predictability of ENSO. First, despite being able to reveal some qualitative information related to the predictability, such as the spring prediction barrier, it remains unclear what the exact upper limit of the forecast skill is. Second, ENSO is intrinsically a chaotic (or turbulent) system, which means quantifying the forecast uncertainty is indispensable. However, the deterministic forecast (including the ensemble average) lacks the ability of the uncertainty quantification in the forecast, without which the meaning of the absolute error is less explainable for chaotic systems. Third, computing these time series based skills requires a large number of data points, which often span for a long observational period. Thus, despite the possible ability to indicate certain overall characteristics, it is challenging to adopt these measurements for understanding the predictability of each individual ENSO event.
However, since the information theory based method utilizes the PDF rather than the time series, it can measure all the linear and non-linear statistical dependence between the ensemble prediction and observation, whereas the traditional signal-to-noise ratio based methods mainly measure their linear correlation and therefore underestimates the nonlinear statistical dependence. Therefore, the information theory based method should be better and more thorough in characterizing the 'true' potential predictability limit \citep{tang2018progress}.
Also, this study suggests that if the information gain rapidly decaying to zero, there is no need to run the model forecast beyond that point since the forecast becomes indistinguishable from the climatological distribution. It is notable that, if the information gain is near zero, the ensemble mean forecast is expected to be close to the climatological mean, which often leads to an unskillful forecast in the time series-based sense as well. Thus, the predictability resulting from the information theory naturally provides an upper bound of the prediction skill, which satisfies the general definition of the predictability.

As stated in Section \ref{Sec:Information_Theory}, the model fidelity is so important for adopting the information theory that we must take it into serious consideration. In practice, stochastic parameterizations and statistical closure approximations can be incorporated into the existing models to improve the statistical accuracy \citep{plant2008stochastic, gottwald2013role,berner2017stochastic, christensen2017stochastic}. The additional components can be calibrated by a certain optimization algorithm with the minimization of the error in the PDFs and ACFs being the cost function. This can be easily implemented for at least simple or conceptual models \citep{sapsis2013statistically, chen2014predicting, chen2018conditional}.

This paper presents a first step towards utilizing the information theory for understanding the predictability of the ENSO complexity. One practical task is to utilize the findings from the information theory based framework as the potential guidelines to improve the existing models and forecast methods.
In particular, taking into account the potential predictability corresponding to the dispersion part deserves more emphasis in the model improvement. Another interesting topic is the multi-model forecast, where the information theory can be used for the model selection that allows an optimal combination of different models to predict each type of the ENSO events. In addition, the exact initial conditions are used in the study of this paper. Yet, data assimilation is an essential step towards the practical forecast, which however will introduce additional uncertainty that may weaken the potential predictability. It is thus important to understand how data assimilation affects the predictability limit of the ENSO complexity and whether coupled atmosphere-ocean data assimilation is helpful in extending the predictability.

Note that the framework introduced in this article can be applied to any kind of ENSO prediction models, provided which are able to simulate the realistic ENSO characters. In the follow-up work and from a practical perspective, the ENSO prediction and the predictability study will be conducted in the general circulation models and the intermediate complicated models, since they can provide more comprehensive interpretation of the nature than the conceptual models. However, before this can be done, we should measure the reasonability of this extension since the more complicated models cannot provide the relevant number of ensemble members like the conceptual model. This requirement is crucial to fit the unbiased PDFs and thus adopt the information theory based framework, but is not always easy to be satisfied since most models we are utilizing are too complicated to simulate the required number of ensemble members, e.g., 1000-10000 as mentioned in \citet{buizza2015forecast}.
And this question can only be answered by utilizing the conceptual model and based on the strict definition of the information theory. This is also of one the main scopes of this work. Encouragingly, based on the conclusion of this work, the Gaussian approximation is shown to be accurate enough in computing the information gain, which facilitates the use of more sophisticated models to study the ENSO predictability in the future work. Note that the Gaussian approximation is utilized only to compute the information measurement, which starts from a specific initial condition. It is not directly related to the overall asymmetric features of the ENSO. In fact, the Gaussian approximation only indicates that  the information gain due to the higher order statistics behaves in a similar way as that from the mean and the variance. It does not suggest a linear model with Gaussian statistics is sufficient for the study of the predictability here.

\acknowledgments
The research of X.F. is supported by Guangdong Major Project of Basic and Applied Basic Research (Grant No. 2020B0301030004), the Ministry of Science and Technology of the People's Republic of China (Grant No. 2020YFA0608802) and the National Natural Science Foundation of China (Grant No: 42192564). The research of N.C. is partially funded by the Office of VCRGE at UW-Madison and ONR N00014-21-1-2904.

\datastatement
The monthly ocean temperature and current data were downloaded from GODAS  (https://www.esrl.noaa.gov/psd/data/gridded/data.godas.html). The daily zonal wind data at 850 hPa were downloaded from the NCEP–NCAR reanalysis (https://psl.noaa.gov/data/gridded/data.ncep.reanalysis.html).






\bibliographystyle{ametsoc2014}


\clearpage
\begin{table}[h]
\centering
\caption{Summary of the non-dimensional units and the model parameters.}
\begin{tabular}{|l|c|l|c|}
  \hline
  $[h]$ & $150$ m & $[T]$ & $7.5 ^o$C\\
  $[u]$ & $1.5$ m/s & $[t]$ & $2$ months\\
  $[\tau]$ & $5$ m/s & $d_\tau$ & $2$\\
  $\gamma$ & $0.75$ & $r$ & $0.25$ \\
  $\alpha_1$ & $0.0625$ & $\alpha_2$ & $0.125$ \\
  $b_0$ & $2.5$ & $\mu$ & $0.5$ \\
  $\sigma$ & $0.2I$ & $\lambda$ & $0.1$ \\
  $p(I)$ & $0.25$ in $I\in(0,4)$ & $c_U$ & $0.03$ \\
  $\Phi(x)$ & $\int_b^x(y-m)p(y)dy$ & $\sigma_I(I)$ & $\sqrt{\frac{2}{p(I)}[-\lambda\Phi(I)]}$\\
  $\beta_E$ & 0.15(2-0.2I) & $\beta_u$ & $-0.2\beta_E$ \\
  $\beta_h$ & $-0.4\beta_E$ & $\beta_C$ & $0.8\beta_E$ \\
  $\sigma_u$ & $0.04$ & $\sigma_h$ & $0.02$ \\
  $\sigma_C$ & $0.04$ & $\sigma_E$ & $0$ \\
  \hline
  $\sigma_\tau(T_C,t)$ & \multicolumn{3}{l|}{$0.9 [\tanh(7.5T_C) + 1]\left[1+0.3\cos\left(\frac{2\pi}{6}t+\frac{2\pi}{6}\right)\right]$} \\
  $c_1(T_C,t)$& \multicolumn{3}{l|}{$\left[25\left(T_C+\frac{0.75}{7.5}\right)^2+0.9\right]\left[1+0.3\sin\left(\frac{2\pi}{6}t-\frac{2\pi}{6}\right)\right]$}\\
  $c_2(t)$& \multicolumn{3}{l|}{$1.4\left[1+0.2\sin\left(\frac{2\pi}{6}t+\frac{2\pi}{6}\right)+0.15\sin\left(\frac{2\pi}{3}t+\frac{2\pi}{6}\right)\right]$}\\
  \hline
\end{tabular}\label{Table_Parameters}
\end{table}

\end{document}